\documentclass[pdflatex,sn-basic]{sn-jnl}% Basic Springer Nature Reference Style/Chemistry Reference Style

\usepackage{graphicx}%
\usepackage{multirow}%
\usepackage{amsmath,amssymb,amsfonts}%
\usepackage{amsthm}%
\usepackage{mathrsfs}%
\usepackage[title]{appendix}%
\usepackage{xcolor}%
\usepackage{textcomp}%
\usepackage{manyfoot}%
\usepackage{booktabs}%
\usepackage{algorithm}%
\usepackage{algorithmicx}%
\usepackage{algpseudocode}%
\usepackage{listings}%
\usepackage{natbib}%
\usepackage{mathabx} % \second (arcsecond)
\usepackage{anyfontsize} % to avoid error message Font shape `U/rsfs/m/n' in size <5.52061> not available
\usepackage{dirtytalk}% quotes
\usepackage{subcaption}% side by side figures
\usepackage{makecell}
\begin{document}

\title[PLATO input catalogs for technical calibration and fine guidance]{\centering PLATO input catalogs for\\ technical calibration and fine guidance}


\author*[1]{\fnm{Ren{\'e}} \sur{Heller}}\email{heller@mps.mpg.de}

\author[1]{\fnm{Chen} \sur{Jiang}}

\author[2]{\fnm{Paz} \sur{Bluhm}}

\author[3,4]{\fnm{Valentina} \sur{Granata}}

\author[5]{\fnm{Juan} \sur{Cabrera}}

\author[5]{\fnm{Denis} \sur{Grie{\ss}bach}}

\author[5]{\fnm{Carsten} \sur{Paproth}}

\author[5]{\fnm{Szil{\'a}rd} \sur{Csizmadia}}
 
\author[5]{\fnm{Philipp} \sur{Eigm{\"u}ller}}

\author[6,7]{\fnm{Paola Maria} \sur{Marrese}}

\author[6,7]{\fnm{Silvia} \sur{Marinoni}}

\author[8]{\fnm{R{\'e}za} \sur{Samadi}}

\author[3,4]{\fnm{Giampaolo} \sur{Piotto}}

\author[9]{\fnm{Marco} \sur{Montalto}}

\author[1]{\fnm{Martin} \sur{Schäfer}}

\author[1]{\fnm{Cilia} \sur{Damiani}}

%\author[10]{\fnm{Magali} \sur{\Unconfirmed{Deleuil}}}

\author[10]{\fnm{Nicholas} \sur{Walton}}

%\author[10]{\fnm{Thierry} \sur{\Unconfirmed{Appourchaux}}}

\author[1]{\fnm{Christoph} \sur{Rauterberg}}

\author[1]{\fnm{Matthias} \sur{Ammler-von Eiff}}

\author[1]{\fnm{Aaron~C.} \sur{Birch}}

\author[1]{\fnm{Laurent} \sur{Gizon}}

%\author[n]{\fnm{first name} \sur{last name}}

\affil*[1]{\orgdiv{Solar and Stellar Interiors Department}, \orgname{Max Planck Institute for Solar System Research}, \orgaddress{\street{Justus-von-Liebig-Weg 3}, \city{G{\"o}ttingen}, \postcode{37077}, \country{Germany}}\vspace{0.2cm}}

\affil[2]{\orgdiv{Fachbereich Geowissenschaften, Institut für Geologische}, \orgname{Freie Universität Berlin}, \orgaddress{\street{Malteserstr. 74-100}, \city{Berlin}, \postcode{12249}, \country{Germany}}\vspace{0.2cm}}

\affil[3]{\orgdiv{Centro di Ateneo di Studi e Attività Spaziali ``Giuseppe Colombo''}, \orgname{Universit{\`a} degli Studi di Padova}, \orgaddress{\street{Via Venezia 1}, \city{Padova}, \postcode{35131}, \country{Italy}}\vspace{0.2cm}}

\affil[4]{\orgdiv{Osservatorio Astronomico di Padova}, \orgname{Istituto Nazionale di Astrofisica},  \orgaddress{\street{Vicolo dell’Osservatorio 5}, \city{Padova}, \postcode{35122}, \country{Italy}}\vspace{0.2cm}}

\affil[5]{\orgdiv{Institute of Space Research}, \orgname{German Aerospace Center}, \orgaddress{\street{Rutherfordstr. 2}, \city{Berlin}, \postcode{12489}, \country{Germany}}\vspace{0.2cm}}

\affil[6]{\orgname{Italian National Institute of Astrophysics}, \orgaddress{\street{Via Frascati 33}, \city{Monte Porzio Catone}, \postcode{00040}, \state{Rome}, \country{Italy}}\vspace{0.2cm}}

\affil[7]{\orgdiv{Space Science Data Center}, \orgname{Agenzia Spaziale Italiana}, \orgaddress{\street{Via del Politecnico, sns}, \city{Rome}, \postcode{00133}, \country{Italy}}\vspace{0.2cm}}

\affil[8]{\orgname{LIRA, Observatoire de Paris, Université PSL, Sorbonne Université, Université Paris Cité, CY Cergy Paris Université, CNRS}, \orgaddress{\city{Meudon}, \postcode{92190}, \country{France}}\vspace{0.2cm}}

%\affil[7]{\orgdiv{Department of Physics and Astronomy ``Galileo Galilei''}, \orgname{University of Padua}, \orgaddress{\street{Vicolo dell’Osservatorio 3}, \city{Padova}, \postcode{35122}, \country{Italy}}\vspace{0.2cm}}

\affil[9]{\orgdiv{Catania Astrophysical Observatory}, \orgname{Italian National Institute of Astrophysics}, \orgaddress{\street{Via S. Sofia 78}, \city{Catania}, \postcode{95123}, \country{Italy}}\vspace{0.2cm}}

%\affil[10]{\orgdiv{Laboratoire d’Astrophysique de Marseille}, \orgname{CNRS, CNES, Aix Marseille University}, \orgaddress{\street{38, rue Frédéric Joliot-Curie}, \city{Marseille}, \postcode{13388}, \country{France}}\vspace{0.2cm}}

\affil[10]{\orgdiv{Institute of Astronomy}, \orgname{University of Cambridge}, \orgaddress{\street{Madingley Rd}, \city{Cambridge}, \postcode{CB3 0HA}, \country{United Kingdom}}\vspace{0.2cm}}

%\affil[10]{\orgdiv{Institut d'Astrophysique Spatiale}, \orgname{University of Paris-Saclay}, \orgaddress{\city{Orsay}, \country{France}}\vspace{0.2cm}}

%\affil[n]{\orgdiv{}, \orgname{}, \orgaddress{\street{}, \city{}, \postcode{}, \state{State}, \country{}}\vspace{0.2cm}}


\abstract{A few weeks after launch, the PLATO spacecraft is expected to start its payload commissioning, which will be completed within the first three months of the mission. This phase includes the in-orbit verification, calibration, and configuration of the instrument prior to nominal science operations. During this mission-critical period, and again later during regular spacecraft rotations and re-pointings, a set of reference stars is required to complete various calibration steps. This set, referred to as the calibration PLATO Input Catalog (cPIC), is part of the PIC. The cPIC comprises various stellar samples, each serving a dedicated technical calibration purpose, and it contains 71\,671 unique stellar targets across PLATO's entire field of view (FoV). Once the spacecraft commences science observations, the on-board Fine Guidance System (FGS) will rely on a small set of guide stars. These stars must be particularly bright and will be observed with the two fast cameras, which cover only a smaller central region of PLATO's FoV. This target list, referred to as the fine-guidance PLATO Input Catalog (fgPIC), contains 2640 unique targets, of which about 30 are used by the FGS at any given time. In this paper, we present the selection criteria for both the cPIC and the fgPIC, and asses their impact on the construction of these calibration catalogs for PLATO.
}


\maketitle

%-------------------------------------------------------------------------------
\section{Introduction}
\label{section:introduction}

The main science objective of the PLATO Mission is the detection of small transiting planets (radii $< 2\,R_\oplus$) in the habitable zones around bright ($m_V \leq 11$) Sun-like stars \citep[spectral types F5–K7;][]{2025ExA....59...26R}. Around a dozen of these Earth-like planets have been predicted to be found with PLATO \citep{2022A&A...665A..11H}, in addition to thousands of other types of exoplanets \citep{2023A&A...677A.133M,2024A&A...692A.150B}. Moreover, PLATO will provide the data for extensive studies of stellar physics via asteroseismology \citep{2017ApJ...835..172L,2019LRSP...16....4G}, gyrochronology \citep{1972ApJ...171..565S,2003ApJ...586..464B,2007ApJ...669.1167B}, activity cycle studies \citep{2021A&A...653A.146G}, measurements of stellar rotation periods \citep{2013A&A...557L..10N,2026arXiv260305586B} etc. with the aim to estimate the bulk physical properties and the ages of solar-type stars.

Many of the mission's technical details that will enable these science goals have been published elsewhere, for example about the spacecraft's multi-telescope approach \citep{2021SPIE11852E..4YP,2022SPIE12180E..4MP}, its fast front end electronics \citep{2022SPIE12180E..1EK}, signal and noise performance \citep{2024ExA....58....1B}, photometry extraction \citep{2019A&A...627A..71M}, and CCD end-to-end simulations \citep{2014A&A...566A..92M,2024A&A...681A..18J}. We refer the reader to these dedicated studies for more details.

In this paper, we present the calibration PLATO Input Catalog \citep[cPIC;][]{cPIC2026} and the fine-guidance PLATO Input Catalog \citep[fgPIC;][]{fgPIC2026} as the two technical calibration input catalogs used by the PLATO Mission. We focus on the selection criteria adopted for the calibration reference stars and on the roles of the cPIC and fgPIC within the overall mission calibration strategy. The implementation of the cPIC and the fgPIC into the PLATO Input Catalog (PIC) is described by \citet{Marrese2026}.

The PIC itself consists of several sub-catalogs (subPICs), including the science target catalog \citep[tPIC;][]{Montalto2026}, the cPIC, the fgPIC, and the science calibration and validation catalog \citep[scvPIC;][]{scvPIC2026,Zwintz2026}. The tPIC contains the main science targets of the mission, while the cPIC is specifically defined to support spacecraft and instrument and performance calibration. The fgPIC provides the reference stars used by the Fine Guidance System (FGS) to ensure accurate spacecraft attitude determination. The scvPIC contains six main stellar samples to calibrate various science cases for PLATO \citep{Zwintz2026}.

%The PIC contains several target samples used for different purposes within the mission. 
An early version of the PIC, referred to as ``asPIC1.1'', included all stars from the second data release of the {\it Gaia} mission \citep[Gaia\,DR2;][]{2018A&A...616A...1G} with spectral types FGK and $V \leq 13$ as well as all M dwarfs and subgiant stars with $V \leq 16$ \citep{2021A&A...653A..98M}. The four sub-catalogs were not fully defined at the time. The latest version of the PIC, referred to as PIC2.2.0.1, was delivered on February 16, 2026, from the PLATO Data Center as part of the PLATO Mission Consortium to the European Space Agency (ESA). On April 7, it is planned that ESA will publish the PIC2.2.0.1 as part of the first call for proposals. The targets proposed by the PLATO guest observers will then pass several validation steps, some of which will also flag them for overlap with any of the targets in the four subPICs.  

The PIC2.2.0.1 contains the cPIC and fgPIC versions described in this paper. Different from the all-sky asPIC1.1, the PIC2.2.0.1 and its subPICs are limited to the first PLATO field \citep[LOPS2;][]{2022A&A...658A..31N,2025A&A...694A.313N}\footnote{An animated version of LOPS2 using the PIC2.2.0.1 is available at \href{https://youtu.be/rU3fas8NFHY}{https://youtu.be/rU3fas8NFHY}.} and to the core program of the mission.

All four subPICs are constructed by design from a table of over 100\,000\,000 targets located in LOPS2 that was extracted from the third data release of the \textit{Gaia} mission \citep[Gaia\,DR3;][]{2016A&A...595A...1G,2023A&A...674A...1G}. This table is called the common list of contaminants of the PIC, and it is shared within the PIC team of the PLATO Data Center \citep{Marrese2026}. The term `contaminants' here is meant to include any sources that could potentially add unwanted photometric flux into the extracted light curves of an adjacent PLATO target. As a consequence, the PLATO targets are considered potential contaminants themselves, which is why they are also present in the common list of PIC contaminants. In the following, we describe the selection of targets for the cPIC and for the fgPIC from this common list of PIC contaminants.

%In addition to the science targets catalog (tPIC), specific subsets of the PIC are defined to support calibration and guidance operations. Among these are the calibration PLATO input catalog (cPIC) and the fine-guidance catalog (fgPIC).  

%The cPIC and the fgPIC are sub catalogs of the PLATO Input Catalog (PIC), therefore internally referred to as ``subPICs'', and they are both needed for different technical calibration steps of the PLATO spacecraft. There are two more subPICs, one of which is the target PIC (tPIC), and the other one being the science calibration and validation PIC (scvPIC). The scvPIC serves the scientific validation and is described in a companion paper \citep{Zwintz2026}.

\subsection{The calibration PLATO Input Catalog (cPIC)}

Achieving PLATO's ambitious science goals requires an accurate calibration and characterization of the PLATO payload throughout its operational lifetime. In particular, several calibration activities must be performed \citep{2019A&A...624A.117S,2025arXiv251022092M} to ensure the accurate determination of the instrumental and observational parameters that affect the quality of the photometric measurements. These calibration activities rely on the identification and observation of suitable reference stars distributed across the field of view (FoV) of the PLATO cameras. These stellar reference samples are referred to as R1, R2, R3, R4, and R5 samples %\citep{cPICreq2024} 
and they are collected in the cPIC.

%The cPIC contains a set of reference targets used to perform several calibration activities during the commissioning phase and subsequent calibration periods of the mission%\citep{cPICreq2024}
%. There is a total of seven reference samples (R1-R7).

Among these, the R1, R2, R3, R4, and R5 samples correspond to stellar targets, while R6 and R7 comprise background windows and offset windows used for calibration purposes, respectively. Since R6 and R7 do not contain any stellar targets, they are not included in the cPIC or in the PIC and are therefore not considered in this document.

The stellar reference samples support different calibration tasks across the FoV of the PLATO cameras (see Table~\ref{tab:ReferenceStars}). The R1 sample is primarily used for spacecraft attitude determination and calibration of the Image Geometry Model (IGM), for which there is one per camera. The R3 sample is used for the determination of the optimal focus position of the cameras, while the R4 sample supports the calibration of the photometric throughput. The stellar reference samples R1, R3 and R4 consist of the same targets and are thus collectively referred to as the the R$i$ samples.
%There will be an overlap between the tPIC and the cPIC which will allow to have continuous monitoring of calibration parameters in nominal operations.

%%%% Whoever wrote this previous sentence, could you please contact me (René) if you would like to have this statement in the paper? I'm sure we will find a way. For now, it felt very much misplaced, when I re-read this passage.

The R2 sample contains stars that will be used during the five-hour microscanning sequence. Microscanning will be part of the spacecraft commissioning phase a few weeks after launch and it will be performed at the beginning of each quarter during nominal science operations. This procedure will aid to measure the point spread function, which is spatially variable across the PLATO field \citep{2019A&A...624A.117S}. 
%We will also perform microscanning sequences during nominal science operations at the beginning of each quarter. 
Some of the R2 selection criteria differ from those applied to the other stellar reference samples because R2 stars are specifically used for the microscanning calibration activity. The R5 sample is used for the determination of the normal-camera (N-CAM) outlier detection threshold in light curves. In practice the R5 stars follow the same selection criteria as the R2 sample for both the N-CAMs and the fast cameras (F-CAMs).\footnote{However, since the F-CAMs only acquire imagettes and not light curves, targets flagged as R5 for the F-CAMs may not necessarily be used for this purpose.}

\renewcommand{\arraystretch}{1.2}
\begin{table}[]
\caption{Reference stellar samples of the cPIC.}
\begin{tabular}{p{1.9cm}|p{9.7cm}}
{\bf Reference star sample} & {\bf Calibration purpose}    \\
\hline
R1                    & Estimation of the telescope attitude and the calibration of the per-camera image geometry model using reference stars distributed across the FoV. \\
\hline
R2                    & Estimation of the point spread function using imagettes obtained during the microscanning sequences. These stars are moderately bright (magnitude $P < 12$, see Sect.~\ref{sec:Pmag}). The R2 sample shall necessarily include all sample P1 stars in addition to other objects that are not included in the P2 sample. \\
\hline
% *?*? \RH{Is this double line a mistake or intentional?}
%\hline
R3                    & Determination of the optimal focus position of the PLATO cameras by evaluating the image quality across the FoV. It shall contain the same stars as the R1 sample. \\
\hline
R4                    & Determination of the photometric throughput of the PLATO cameras using reference stars distributed across the FoV. It shall contain the same stars as the R1 sample.\\
\hline
R5                    & Determination of the N-CAM outlier detection threshold for light curves. For the first PIC of the LOPS2 field (PIC2.2.0.1), the R5 sample has the same selection criteria as the R2 sample. The selection criteria might be adapted based on experience with mission data.
\end{tabular}
\footnotetext{{\bf Notes.} The R2 and R5 samples serve different purposes, but they currently have the same selection criteria. The stellar reference samples R1, R3, and R4 are collectively referred to as R$i$.}
\label{tab:ReferenceStars}
\end{table}

%The calibration data acquired for the fgPIC (\PB{Should this refer to the cPIC rather than the fgPIC}) are essentially of two types:
%\begin{itemize}
%\item Full-frame CCD images: four per camera.
%\item Imagettes: about 14,000 targets with apparent PLATO magnitudes ranging %approximately from 8.5 to 12. 
%\end{itemize}

The calibration observations of the cPIC targets will provide two main types of data products:

\begin{itemize}
\item full-frame CCD images (four per camera)
\item imagettes extracted for approximately about 14\,000 targets with apparent PLATO magnitudes $P$ (see Sect.~\ref{sec:Pmag}) ranging approximately from 8.5 to 12.
\end{itemize}

%The cPIC reference stars will also be used to identify the observed FoV during calibration observations. Bright stars detected in the full-frame CCD images can be matched with stars from external all-sky catalogs such as Hipparcos and Tycho. This comparison will allow the observed field to be uniquely identified and will provide the reference frame needed to identify the corresponding cPIC targets on these images. Once the field has been identified, the corresponding cPIC targets will be located in the images and associated with their respective reference samples, enabling the calibration measurements required for instrument characterization. \RH{($\Leftarrow$ Paz, please help to avoid redundancies and boil it down to the point.)}

%\PB{During calibration observations, the observed FoV is identified from bright stars, non-saturated stars detected in the full-frame CCD images through comparison with a stellar catalog. Once the field has been identified, the cPIC reference stars can be located in the images and used for the corresponding calibration activities.}

During calibration observations, the observed FoV will be identified using a separate sample of bright ($m_V < 8$), non-saturated stars in the full-frame CCD images through comparison with a stellar catalog. Once the field has been identified, the cPIC reference stars can be located in the images and used for the corresponding calibration activities.

\subsection{The fine-guidance PLATO Input Catalog (fgPIC)}

To achieve the very ambitious science goals, the spacecraft also needs to maintain a very high pointing accuracy of $0.025''$ \citep{2021SPIE11852E..3HG}. Translated into the pixel scale, given that one pixel covers $15''$ and the pixel edge length is $18\,{\mu}$m, this means a noise equivalent angle of 1/600 pixel. In other words, the pointing shall always be more accurate than $0.03\,{\mu}{\rm m} = 3$\,nm on the CCD.

The on-board FGS of PLATO sits at the core of this task. It will be fed with data from the two F-CAMs at a cadence of 2.5\,s, the latter of which was chosen to satisfy the requirement that stars with apparent magnitudes $< 8.2$ are not to be saturated. The optical design of the two F-CAMs is identical, but the filter response functions differ between the blue F-CAM, being sensitive between 500\,nm and 662\,nm, and the red F-CAM, being sensitive between 662\,nm and 1000\,nm \citep{2024A&A...681A..18J,Cabrera2026,2026arXiv260312750L}. PLATO's FGS algorithm has been optimized to perform all the necessary computations using the F-CAM data of about 30 bright, photometrically stable stars within 300\,ms \citep{2021SPIE11852E..3HG}.

One important effect on the pointing stability of the PLATO spacecraft is the stellar variability, which as already been investigated for the FGS of the Kepler Space Mission \citep{2010Christiansen}. Activity of the fine-guidance stars or of their unseen stellar companions can affect the inferred centroids. With the fine-guidance algorithms trying to correct for these tiny variations, this might induce a systematic effect on the science observations of the target stars as their positions could vary very slightly on the CCD. An in-depth study of such effects of stellar variability on PLATO's pointing stability is given by \citet{2026AJ....171...14B}.

\section{Selection criteria for the cPIC}
\label{sec:cPIC}

\renewcommand{\arraystretch}{1.2}

\begin{table}[]
\caption{Selection criteria for the cPIC targets.}
\begin{tabular}{p{2.09cm}|p{7.1cm}|p{3.05cm}}
{\bf Steps} & {\bf Selection criteria} & {\bf No. of targets} \\
\hline
1. Cone search & The cPIC shall include targets in the LOPS2 field, centered at RA = 95$^\circ$.31043, Dec = -47$^\circ$.88693. & {\bf N-CAMs:} 105\,865\,126 \newline \newline {\bf F-CAMs:} 23\,594\,558 \\
\hline
2. PLATO \newline \hspace*{0.25cm} magnitude \newline \hspace*{0.25cm} range & {\bf N-CAMs:} \newline In the N-CAMs, the cPIC stars shall have the following apparent magnitudes. \newline R$i$: $8.5 < P < 9.5$ \newline R2: $P < 12$ \newline \newline {\bf F-CAMs:} \newline In the F-CAMs, the cPIC stars shall have the following apparent magnitudes. \newline  R$i$: $3.7 < P < 8.2$ \newline R2: $4 < P < 12$ & {\bf N-CAMs:} \newline R$i$: 20\,895 \newline R2: 321\,845 \newline \newline {\bf F-CAMs:} \newline R$i$: 4857 \newline R2: 142\,569 \\
\hline
3. Stellar \newline \hspace*{0.25cm} Color & {\bf N-CAMs:} \newline In the N-CAMs, the cPIC stars shall have the following spectral types. \newline R$i$: F5 to K7 (P1 and P5 Samples) and K7 to M5 (P4 Sample) \newline R2: F5 to K7 (P1,P5) and K7 to M5 (P4) \newline \newline {\bf F-CAMs:} \newline In the F-CAMs, the cPIC stars shall have the following spectral types. \newline R$i$: F5 to K7 (P1 and P5 Samples) and K7 to M5 (P4 Sample) \newline R2: dwarfs and subgiants later than spectral type F5 (P2 Sample) & {\bf N-CAMs:} \newline R$i$: 16\,127 \newline R2: 264\,418 \newline \newline {\bf F-CAMs:} \newline R$i$: 584 \newline R2: 62\,033 \\
\hline
4. Photometric \newline \hspace*{0.25cm} contaminants & The cPIC shall not include targets for which the integrated flux of contaminants within 4 pixels ($60''$) is less than four magnitudes fainter than the calibration target itself.
\newline \newline
Waiver: If there are regions of the sky with equal area of 1 square degree without stars fulfilling this requirement, then this requirement is ignored in these regions. & {\bf N-CAMs:} \newline R$i$: 10,532 \newline R2: 62\,639 \newline \newline {\bf F-CAMs:} \newline R$i$: 547 \newline R2: 17\,800 \newline \newline {\bf Total:} \newline 71\,671 unique targets \\
\hline
5. Accuracy of \newline \hspace*{0.25cm} coordinates & Barycentric Celestial Reference System coordinates shall be known with an accuracy better than 150 mas.  &  No targets rejected. \\
\hline
6. Proper \newline \hspace*{0.25cm} motion & In case a target has no referenced proper motion, it is rejected from the cPIC. If the precision of the proper motion is $> 900$\,mas/year, then the target is rejected from the cPIC. For bright stars with either $P \leq 6$ or $P_{\rm red} \leq 6$ or $P_{\rm blue} \leq 6$ there is a waiver for the precision or availability of the proper motion. These stars are generally not rejected based on proper motion criteria. & No targets rejected. \\
\hline
7. Stellar \newline \hspace*{0.25cm} variability & The cPIC shall flag known eclipsing binaries, visual binaries, long-period variables, flaring stars, and Cepheids. The flags are taken from Gaia\,DR3. &  No targets rejected. %\\
%\hline
%8. Output & The cPIC table is created using columns defined in the data model of the PIC. & No targets rejected.
\end{tabular}
\label{tab:cPIC}
\end{table}

The construction of the cPIC is based on a set of selection criteria that include spatial constraints within the LOPS2 field, magnitude and color limits, photometric contamination thresholds%\citep{cPICreq2024}
, and additional requirements on astrometric accuracy, proper motion, and stellar variability that apply to all PIC targets%\citep{PICreq2024}
. A summary of the selection criteria for the cPIC is given in Table~\ref{tab:cPIC}, where the selection of the R5 targets is covered by the selection of the R2 targets since both samples are initially the same.

%%%%%%%%%%%%%%%%%%%%%%%%%%%%%%%%%%%%%%%%%%%%%%%%%%%55
%subsections 2.1 - 2.8 Added by Paz work in progress...
\subsection{Cone search}
\label{sec:cone_cPIC}

The first step in the selection process for the cPIC consists of identifying candidate targets located within LOPS2. For cPIC targets observed by the N-CAMs, we searched the list of PIC contaminants within a cone with a radius of $30^\circ$ around the LOPS2 center to cover the entire N-CAM FoV, with a calculated radius of approximately $28^\circ$. We add $2^\circ$ of radial margin to account for possible re-pointings of the spacecraft. For cPIC targets seen by the F-CAMs, we searched for targets around the LOPS2 center within a cone of $19^\circ$ radius. This radius covers the entire F-CAM field of view (FoV) of approximately $18^\circ.9$, plus some margin.

\subsection{PLATO magnitude}
\label{sec:Pmag}

The PLATO magnitude is one of the main parameters used in the selection of calibration reference stars. The adopted magnitude ranges are chosen to ensure sufficiently high signal-to-noise ratios while avoiding saturation in the PLATO cameras. The exact magnitude limits depend on the calibration sample considered, since each sample supports a different calibration tasks and therefore requires different observational properties. 

In general, the R$i$ samples observed with the N-CAMs are restricted to relatively bright stars suitable for high-precision measurements, while broader magnitude ranges are adopted for samples requiring a larger number of targets distributed across the field-of-view. The R2 sample, on the other hand, will be used for the microscanning sequence and will require more moderately bright stars that allow precise measurements of the point spread function (PSF) from imagettes without saturating the detector.
%The magnitude ranges adopted for the different cPIC samples are summarized in Table~\ref{tab:cPIC}.

To search for cPIC targets that will be observable by PLATO's 24 N-CAMs, we use the PLATO N-CAM magnitude $P$ \citep{2019A&A...627A..71M, Montalto2026}, whereas for cPIC targets that could be observed with the F-CAMs, we use the PLATO magnitudes in the blue and red F-CAM passbands ($P_{\rm blue}$ and $P_{\rm red}$), respectively \citep{2024A&A...681A..18J,Cabrera2026}.

\subsection{Stellar color}

We use the stellar color as a selection criterion to ensure that the calibration targets have suitable spectral properties for the different calibration tasks. The adopted color ranges are chosen to preferentially select stars with spectral types representative of the PLATO core science targets that will also be used for scientific exploitation.% This ensures that the calibration targets exhibit photometric properties similar to those expected for the science targets. 
%The exact color limits depend on the calibration sample considered and are summarized in Table~\ref{tab:cPIC}.

We use the $G$, $G_{\rm BP}$ and $G_{\rm RP}$ Gaia\,DR3 magnitudes \citep{2010A&A...523A..48J} as an input to the PIC2PARAM code\footnote{The PIC2PARAM code evaluates the expected interstellar extinction for given coordinates on the celestial plane, and then calculates the PLATO magnitude in PLATO's N-CAMs and F-CAMs for stars with input photometry from Gaia DR3, 2MASS, Hipparcos, Tycho or with known Johnson photometry. Further details of the PIC2PARAM code are detailed in \citet{Montalto2024} and \citet{Montalto2026}.} v3.4 \citep{Montalto2024}. PIC2PARAM then generates a caseFlag, of which the following allowed values are relevant to the cPIC targets:

\begin{itemize}
\item caseFlag = 1: FGK dwarfs and subgiants
\item caseFlag = 2: M dwarfs
\item caseFlag = 4: giants
\end{itemize}

\noindent
The caseFlag is saved in the caseFlag column of the cPIC for each target.

\subsection{Photometric contamination}

Photometric contamination from nearby sources can affect the measured flux of calibration targets and degrade the quality of the photometric measurements. To minimize these effects, an additional selection criterion is applied to exclude targets with significant contaminating flux from nearby stars within the photometric aperture.
%The adopted thresholds for the different calibration samples are summarized in Table~\ref{tab:cPIC}.

\begin{figure}[h!]
\centering
\includegraphics[width=0.495\textwidth]{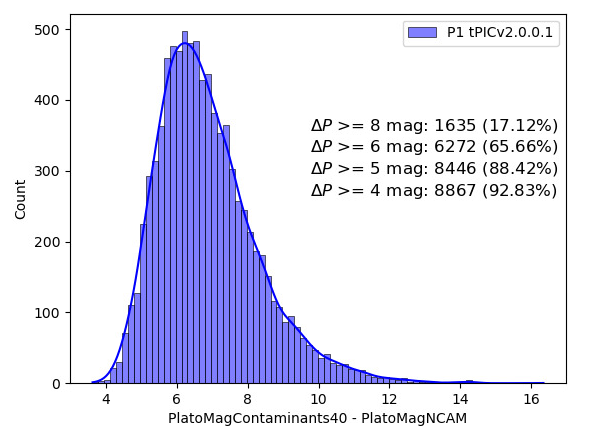}
\includegraphics[width=0.495\textwidth]{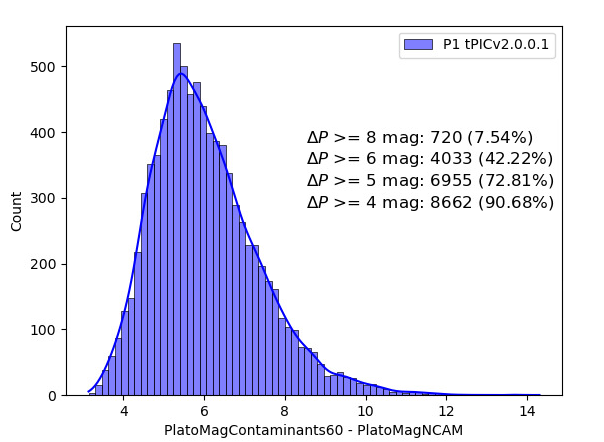}\\
\includegraphics[width=0.495\textwidth]{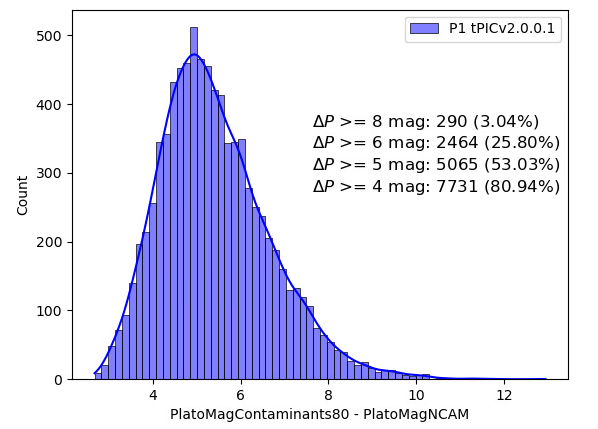}
\includegraphics[width=0.495\textwidth]{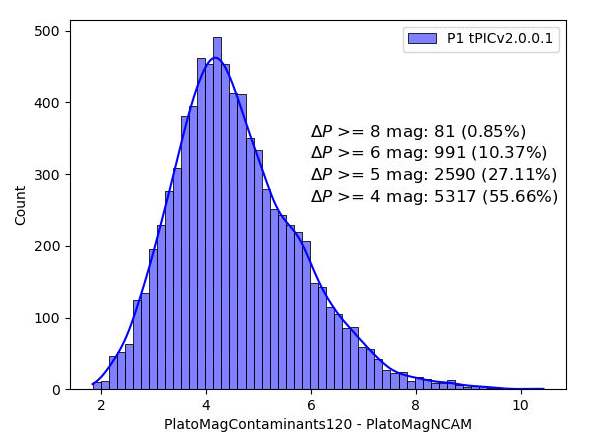}
\caption{Histograms of the PLATO magnitude difference of the P1 Sample targets and the combined magnitude of the contaminants within a given radius of each P1 star. Legends summarize the number and the fraction of P1 stars that would have been selected into the LOPS2 cPIC if the respective ${\Delta}P$ had been chosen. \textit{Top left:} Radius for contaminants: $40''$. \textit{Top right:} Radius for contaminants: $60''$. \textit{Bottom left:} Radius for contaminants: $80''$. \textit{Bottom right:} Radius for contaminants: $120''$.}
\label{fig:P1tocPIC}
\end{figure}

As mentioned in Table~\ref{tab:ReferenceStars}, the R2 sample (and therefore the cPIC) shall include all P1 Sample\footnote{In brief, the P1 Sample ``includes at least 15,000 dwarf and sub-giant stars (types F5 to K7), cumulative over the nominal mission, with $V \leq 11$\,mag and a noise level of $<50$\,ppm in 1\,h'' \citep{2025ExA....59...26R}.} stars. In our initial investigations of the cPIC generation, we learned that the selection for stars with negligible contaminants actually rejects many P1 Sample stars from the cPIC. To better understand the effect of photometric contamination, we computed the contamination of the P1 Sample stars by nearby targets with four different contamination radii ($40''$, $60''$, $80''$, $120''$) around the target and four different cutoff magnitudes ${\Delta}P \in \{4,5,6,8\}$ for the combined contribution of the contaminants. Figure~\ref{fig:P1tocPIC} shows the resulting histograms of the PLATO magnitude difference of the P1 stars from the most up to date PIC version at the time (v2.0.0.1).

These tests demonstrate that a contamination radius of $40''$ (Fig.~\ref{fig:P1tocPIC}, upper left panel) and ${\Delta}P \geq 4$ are required to keep as many P1 sample stars as possible (92.83\,\%) in the cPIC. Increasing the contamination radius by 50\,\% from $40''$ to $60''$ (Fig.~\ref{fig:P1tocPIC}, upper right panel), however, ensures much better isolation of the cPIC stars against photometric contaminants, while still preserving 90.68\,\% of the P1 Sample for ${\Delta}P \geq 4$. Since photometric stability is key to enabling the important inversion of the point spread function and since the effect of the P1 Sample fraction in the cPIC only varies by 2.15\,\% between $40''$ and $60''$ as a contamination radius, we therefore decided to apply a contamination radius of $60''$ and a magnitude difference of ${\Delta}P \geq 4$ between the cPIC targets and the combined contaminant flux.

\begin{figure}[t]
\centering
\includegraphics[width=0.99\textwidth]{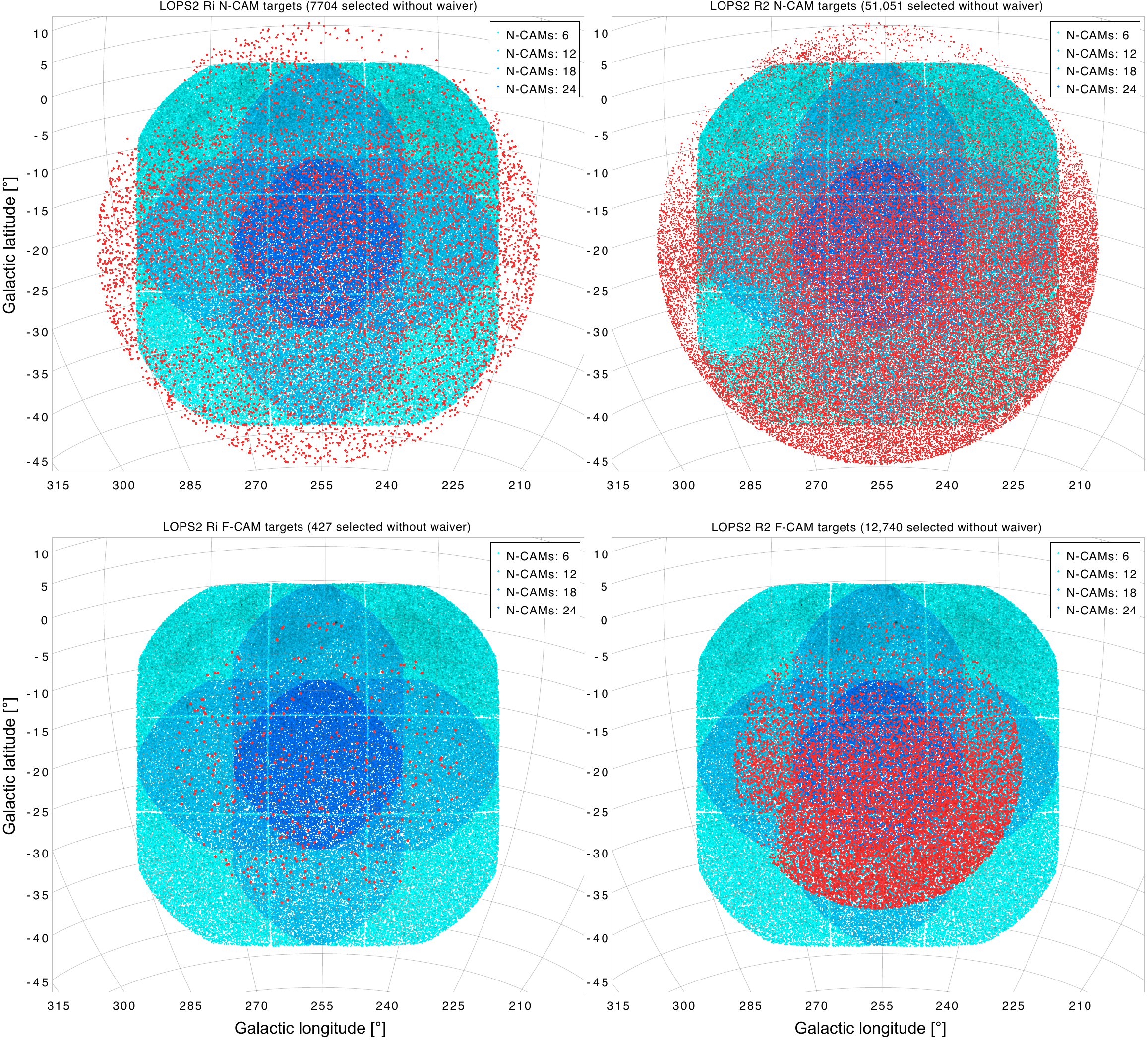}
\caption{Sky plot of the cPIC targets (red dots) that were selected before the waiver for photometric contaminants in crowded fields was introduced. The blue shaded regions are covered by targets from the LOPS2 PIC2.2.0.1. Light to dark blue regions correspond to a coverage by 6, 12, 18, and 24 N-CAMs, respectively. The apparent hole Galactic longitude of $280^\circ$ and Galactic latitude of $-33^\circ$ is caused by stellar crowding in the Large Magellanic Cloud and the resulting rejection of cPIC targets due to photometric contamination. \textit{Top left:} R$i$ N-CAM sample. \textit{Top right:} R2 N-CAM sample. \textit{Bottom left:} R$i$ F-CAM sample. \textit{Bottom right:} R2 F-CAM sample.}
\label{fig:cPICNCAM_NoWaiver}
\end{figure}

\begin{figure}[t]
\centering
\includegraphics[width=0.99\textwidth]{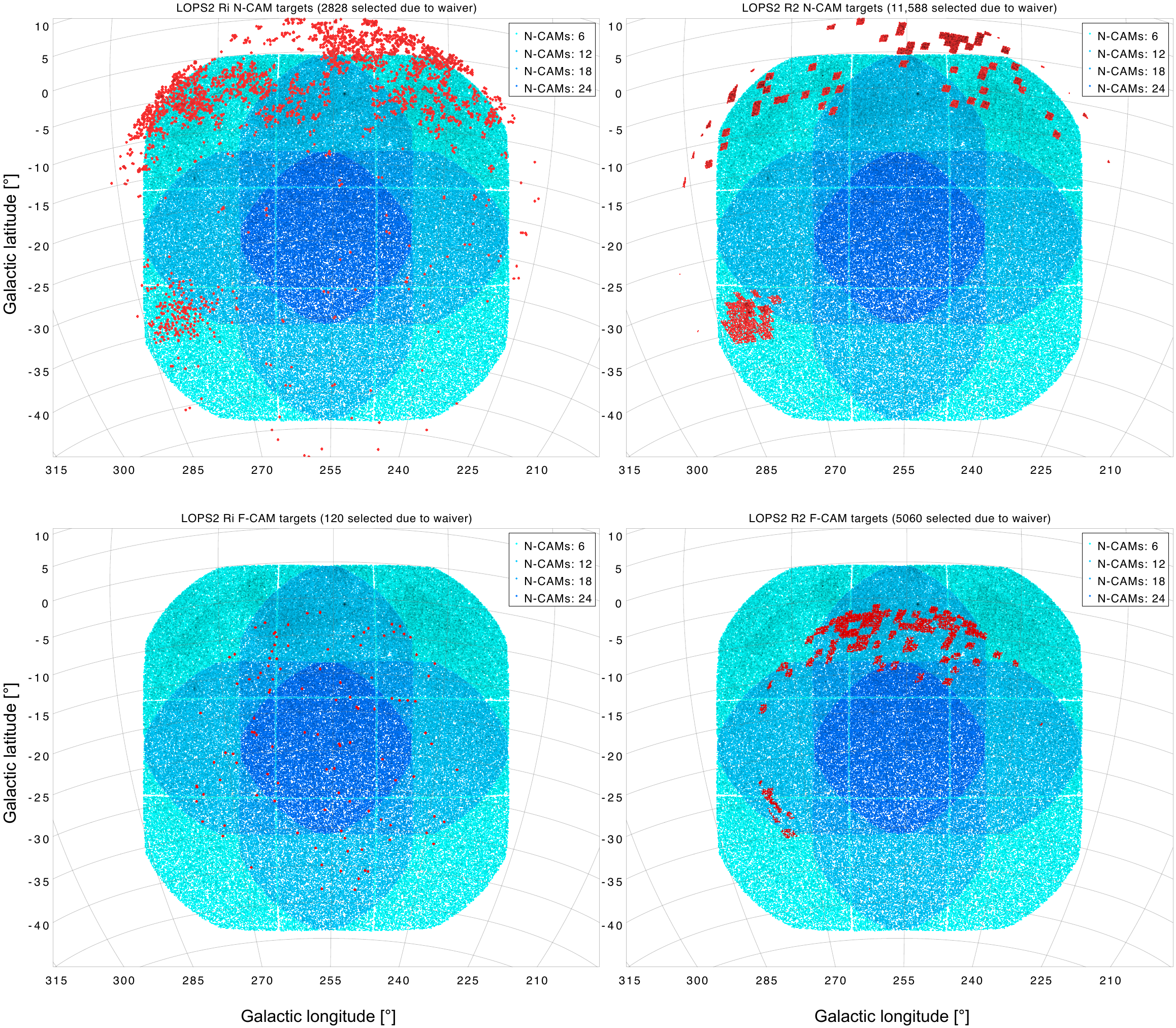}
\caption{Sky plot of the cPIC targets (red dots) that were selected due to the waiver for photometric contaminants in crowded fields.  \textit{Top left:} R$i$ N-CAM sample. \textit{Top right:} R2 N-CAM sample. \textit{Bottom left:} R$i$ F-CAM sample. \textit{Bottom right:} R2 F-CAM sample.}
\label{fig:cPICNCAM_contamination}
\end{figure}

Application of this selection criterion for photometric contamination results in peculiar ``holes'' in the sky distribution of the cPIC target stars, as illustrated in Fig.~\ref{fig:cPICNCAM_NoWaiver}. These empty areas are introduced due to the overwhelming stellar crowding in the Large Magellanic Cloud (near a Galactic longitude of $280^\circ$ and Galactic latitude of $-33^\circ$) and towards the Galactic plane near Galactic latitude of zero. To mitigate this effect, we introduced a waiver for the rejection of photometrically contaminated targets from the cPIC in order to preserve a sufficient number of targets across the full FoV. For any region within the cPIC selection FoV (see Sect.~\ref{sec:cone_cPIC}) that has an area of at least one square degree and that ends up with zero cPIC stars after the selection for contamination, this requirement is ignored. This waiver acts to ``patch'' the holes observed in the sky distribution of cPIC targets. The outcomes are shown in Fig.~\ref{fig:cPICNCAM_contamination}.

A few slightly underpopulated regions in the sky distribution remain in regions where there is just barely a sufficient stellar density of cPIC targets to prevent the waiver from being called. So these remaining open spots are just a little bit smaller than one square degree. On the other hand, spots that are just a little bit larger than one square degree get completely filled with targets due to the waiver, which results in a few relatively overpopulated areas, for example in the region of the Large Magellanic Cloud and near the Galactic plane (see Fig.~\ref{fig:cPICNCAM_contamination}).

Note that the number of targets in the panel titles of Figs.~\ref{fig:cPICNCAM_NoWaiver} and \ref{fig:cPICNCAM_contamination} add up to the total sample counts listed in Table~\ref{tab:cPIC}. As a reading example, after selection step 4 for photometric contaminants, the upper left panel of Figs.~\ref{fig:cPICNCAM_NoWaiver} mentions 7704 targets in the R$i$ sample for the N-CAMs that were selected prior to the application of the waiver. Then the upper left panel of Fig.~\ref{fig:cPICNCAM_contamination} lists another 2828 targets in this sample that were selected due to the waiver. The sum of 10\,532 targets in that sample is listed in the corresponding category in the last column of line four in Table~\ref{tab:cPIC}.

%Added by Paz work in progress...
\subsection{Accuracy of coordinates}
Accurate astrometric positions are required to ensure the reliable identification of the calibration targets and the correct placement of the photometric apertures on the detector. Therefore, an additional constraint on the astrometric accuracy of the target coordinates is applied during the selection process.
%The detailed criteria applied to the proper motion measurements are summarized in Table~\ref{tab:cPIC}.

\subsection{Availability of proper motion}
Proper motion information is required to ensure that the predicted positions of the calibration targets remain accurate over the mission lifetime. Targets without proper motion measurements, or with highly uncertain proper motion estimates, may lead to incorrect target positions and are therefore rejected from the cPIC selection. 
%The detailed criteria applied to the proper motion measurements are summarized in Table~\ref{tab:cPIC}.

\subsection{Stellar variability}
\label{sec:variability}

Stellar variability can affect the stability of the photometric reference targets used for calibration. For this reason, information on known stellar variability is collected for all selected targets, though none of the cPIC targets is initially rejected due to their known photometric activity. We store for each target the phot\_variable\_flag from Gaia\,DR3 in the variable fgPICcPICvariabilityFlag:

%. In particular, known eclipsing binaries, visual binaries, long-period variables, flaring stars, and Cepheids are flagged using information from Gaia\,\,DR3.
%The adopted variability flags and their usage in the cPIC selection are summarized in Table~\ref{tab:cPIC}.

\begin{itemize}
\item 0: non-variable
\item 1: variable
\item 2: not available
\end{itemize}

\section{The cPIC}
\label{sec:output_cPIC}

The final output of the selection process is the cPIC, version LOPS2cPICtarget2.2.0.3 of which is illustrated in this paper and which was merged into the PIC2.2.0.1 to be used for the first PLATO call for proposals. The cPIC table is created using columns defined in the data model of the cPIC \citep{Marrese2026}, and it follows mostly the data model defined for the PIC. The catalog contains the relevant stellar identifiers, astrometric parameters, photometric information, and flags describing the properties of the selected targets. As a so-called Intermediate Data Product, it is not made public as a separate table. But all the information about the cPIC targets is contained in the PIC and encoded in the cPICsourceFlag.

\renewcommand{\arraystretch}{1.2}
\begin{table}[h!]
\caption{Definition of the bitmask values of the cPICsourceFlag}
\begin{tabular}{p{1.58cm}|p{1.05cm}|p{2.99cm}}
{\bf Bitmask} & {\bf Integer} & {\bf Calibration purpose}    \\
\hline
0000000000 & 0 & reset all bits \\
0000000001 & 1 & R1 star for F-CAM \\
0000000010 & 2 & R2 star for F-CAM \\
0000000100 & 4 & R3 star for F-CAM \\
0000001000 & 8 & R4 star for F-CAM \\
0000010000 & 16 & R5 star for F-CAM \\
0000100000 & 32 & R1 star for N-CAM \\
0001000000 & 64 & R2 star for N-CAM \\
0010000000 & 128 & R3 star for N-CAM \\
0100000000 & 256 & R4 star for N-CAM \\
1000000000 & 512 & R5 star for N-CAM
\end{tabular}
\label{tab:cPICsourceFlag}
\end{table}

The cPICsourceFlag records the membership of the targets in any of the reference star samples. It is stored as an integer value in the cPIC (and in the PIC), which is computed from the binary bitmask values shown in Table~\ref{tab:cPICsourceFlag}. The cPIC contains a total of 71\,671 unique targets, with more details about the membership in the R$i$ or R2 samples and the observability in the N-CAMs or F-CAMs in Table~\ref{tab:cPIC} (see selection step 4 therein). Curiously, we find that the following nine targets turn out to be present in all those samples, that is, their cPICsourceFlag = 1023. Three targets that are also in the fgPIC are labeled with an asterisk (\textsuperscript{*}), and one target from the scvPIC1b sample of astrometric binaries \citep{Zwintz2026} is indicated with a dagger (\textsuperscript{\textdagger})

\begin{itemize}
\item PIC\,2971364000451 (Gaia\,DR3\,4757524489124658048)\textsuperscript{\textdagger}
\item PIC\,2907352000055 (Gaia\,DR3\,5495052596695570816)\textsuperscript{*}
\item PIC\,2899054000140 (Gaia\,DR3\,5486695861648434304)
\item PIC\,2886247000110 (Gaia\,DR3\,5486981524216356608)
\item PIC\,2785732000054 (Gaia\,DR3\,4793601557272771968)\textsuperscript{*}
\item PIC\,2750997000060 (Gaia\,DR3\,4794632903476180096)\textsuperscript{*}
\item PIC\,2712773000006 (Gaia\,DR3\,4799007688445126016)
\item PIC\,2639166000058 (Gaia\,DR3\,4805806449875760384)
\item PIC\,2871599000151 (Gaia\,DR3\,5497072086021346688)
\end{itemize}

\begin{figure}[t]
\centering
\includegraphics[width=0.95\textwidth]{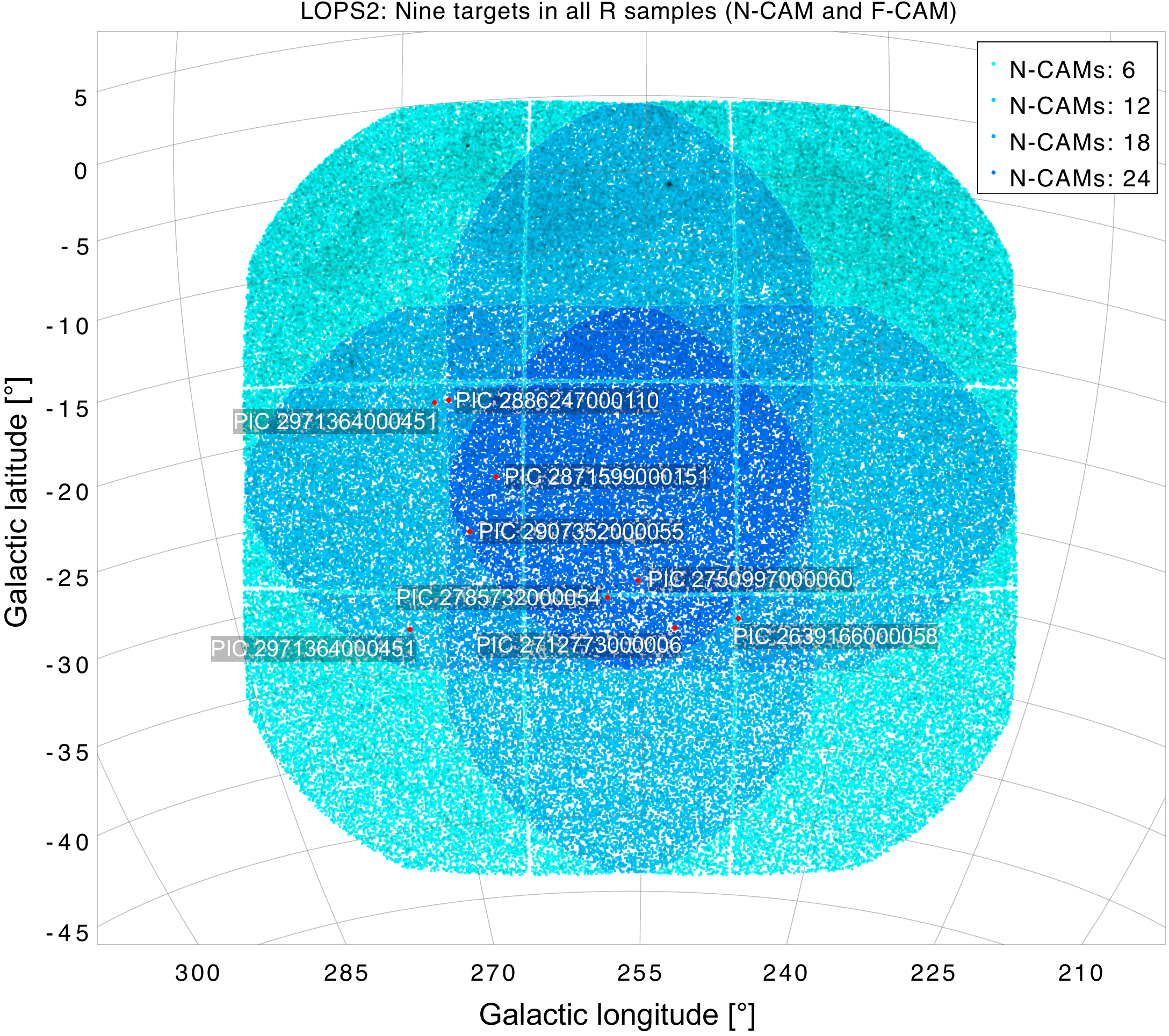}
\caption{Sky plot of nine cPIC targets that are present in all stellar reference samples. Their PIC names are indicated with labels.}
\label{fig:cPICstarsInAllRSamples}
\end{figure}

\noindent
They all have PLATO N-CAM magnitudes $8.51~{\leq}~P~{\leq}~8.72$, PLATO F-CAM (blue) magnitudes $8.80~{\leq}~P_{\rm blue}~{\leq}~9.46$, PLATO F-CAM (red) magnitudes $8.06~{\leq}~P_{\rm red}~{\leq}~8.20$, and by virtue of the selection criteria, they are in the FoV of the F-CAMs. Their positions in the LOPS2 field are shown in Fig.~\ref{fig:cPICstarsInAllRSamples}. 

%\PB{Hi Ren{\'e}, Do you think it would be useful to include a small table summarizing the number of selected targets? For example, the number of targets per sample (R1–R5), per camera type, or the total distribution across the FoV.?}

%\RH{Hey Paz! Some information about the number of targets in the Rx samples is given in Table~\ref{tab:cPIC}. If you would like to show more info, please go ahead and you could add it as a new column ``Sample count'' (just a proposal) in Table~\ref{tab:cPICsourceFlag}.}

%\PB{ok, I will think about it}

\section{Selection criteria for the fgPIC}
\label{sec:selection_fgPIC}

The construction of the fgPIC is based on a set of selection criteria that include spatial constraints within the LOPS2 field, magnitude and color limits, photometric contamination thresholds%\citep{fgPICreq2023}
, and additional requirements on astrometric accuracy, proper motion, and stellar variability that apply to all PIC targets%\citep{PICreq2024}
. A summary of the selection criteria for the fgPIC is given in Table~\ref{tab:fgPIC}.

\subsection{Cone search}

The first step in the selection process for the fgPIC consists of identifying candidate targets located within the FoV of the F-CAMs. We searched the list of PIC contaminants within a cone with a radius of $19^\circ$ around the LOPS2 center to cover the entire F-CAM FoV, with a calculated radius of approximately $18^\circ.9$, plus some margin.

\subsection{PLATO magnitude}

We then reduced the list of selected targets in the F-CAM viewing zone of LOPS2 to those targets that fulfill at least one of the constraints on either the PLATO blue magnitude ($P_{\rm blue}$) or the PLATO red magnitude ($P_{\rm red}$). Details are given in Table~\ref{tab:fgPIC}, selection step 2. Depending on which condition is fulfilled, we updated a binary bitmask called fgPICsourceFlag accordingly:

\begin{itemize}
\item 000000 = 0 : not in F-CAM blue and not in F-CAM red
\item 000001 = 1 : in F-CAM blue
\item 000010 = 2 : in F-CAM red
\item 000011 = 3 : in F-CAM blue and in F-CAM red,
\end{itemize}

\noindent
where the four binary digits on the left-hand side encode additional information about the target membership in the tPIC, fgPIC, cPIC, and scvPIC, respectively.

\renewcommand{\arraystretch}{1.2}

\begin{table}[]
\caption{Selection criteria for the fgPIC targets.}
\begin{tabular}{p{2.09cm}|p{7.1cm}|p{2.52cm}}
{\bf Steps} & {\bf Selection criteria} & {\bf No. of targets} \\
\hline
1. Cone search & The fgPIC shall include targets in a $19^\circ$ angle cone around the center of the LOPS2 field at (RA = 95$^\circ$.31043, Dec = -47$^\circ$.88693). & {\bf Total:} 23\,594\,559 \\
\hline
2. PLATO \newline \hspace*{0.25cm} magnitude \newline \hspace*{0.25cm} range & The fgPIC stars shall have the following magnitudes. \newline \newline {\bf F-CAM (blue):} \newline $4.5 < P_{\rm blue} < 8.5$ \newline \newline {\bf F-CAM (red):} \newline $4.5 < P_{\rm red} < 8.5$ & {\bf F-CAM (blue):} \newline 3126 \newline \newline {\bf F-CAM (red):} \newline 6445 \\
\hline
3. Photometric \newline \hspace*{0.25cm} contaminants & The fgPIC shall not include targets for which the integrated flux of contaminants within 4 pixels ($60''$) is less than 6 magnitudes fainter than the calibration target itself. & {\bf F-CAM (blue):} \newline 1801 \newline \newline {\bf F-CAM (red):} \newline 2174 \newline \newline {\bf Total unique:} \newline 2640 \\
\hline
4. Stellar \newline \hspace*{0.25cm} variability & The fgPIC shall flag known eclipsing binaries, visual binaries, long-period variables, flaring stars, and Cepheids based on flags from Gaia\,DR3. &  No targets rejected.% \\
%\hline
%5. Output & The fgPIC table is created using columns defined in the data model of the fgPIC. & No targets rejected.
\end{tabular}
\label{tab:fgPIC}
\end{table}

\subsection{Photometric contamination}
\label{sec:PhotContfgPIC}

Targets that qualify for the fgPIC by means of their position in the F-CAM viewing zone and as per their PLATO magnitude can potentially be contaminated by the flux from other nearby stars. In principle, such contaminants can be in the foreground or in the background of the target, though this aspect is irrelevant in our context. Any targets that suffer from substantial photometric contamination from nearby sources must be rejected from the fgPIC because otherwise effects like astrophysical activity (eclipses, flares, high-energy events etc.) could affect the target's center of brightness and therefore sabotage the FGS. The constraints on flux contamination is listed in Table~\ref{tab:fgPIC}, selection step 3.

\begin{figure}[t]
\centering
\includegraphics[width=0.95\textwidth]{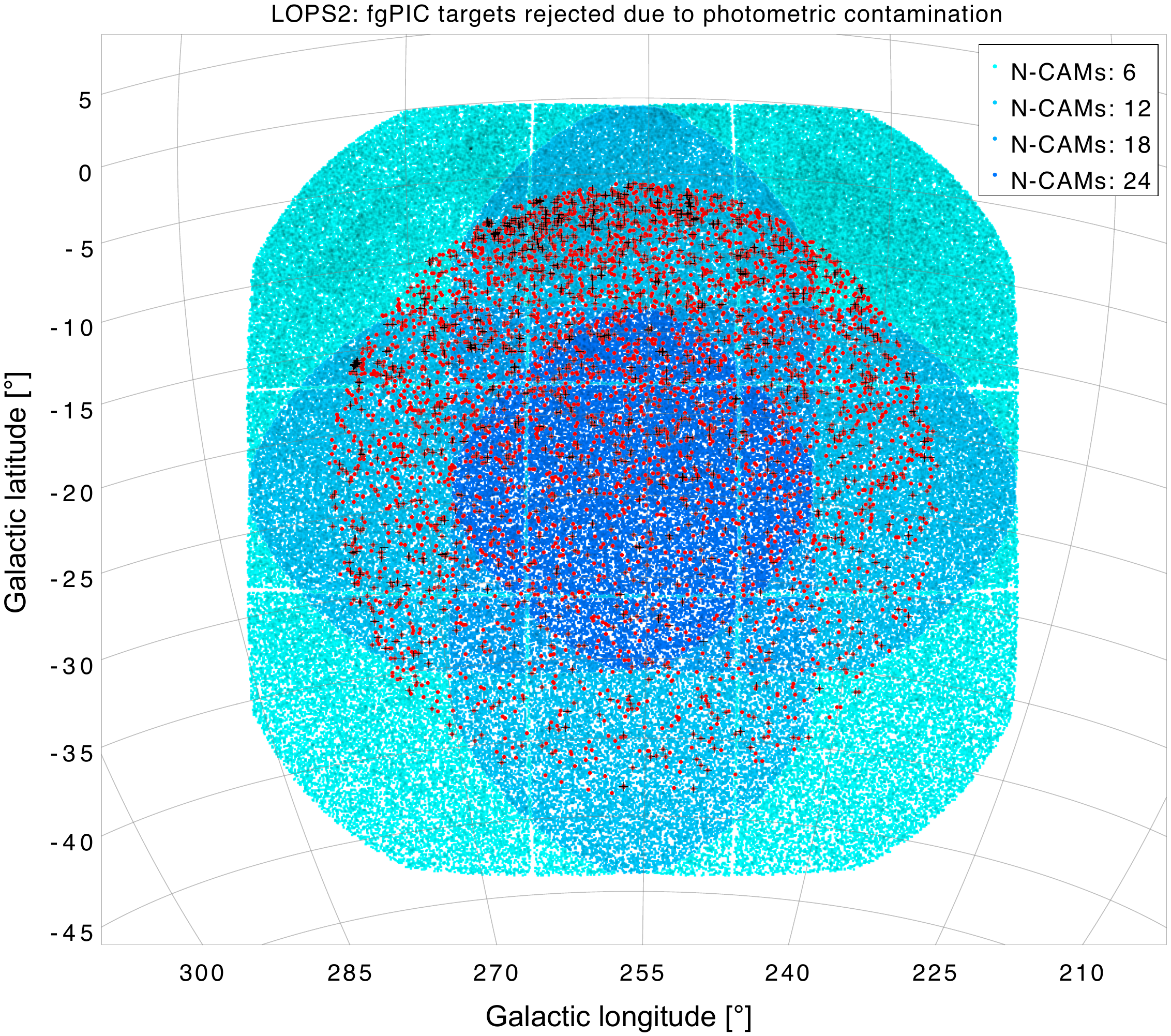}
\caption{Sky plot of the fgPIC targets that were rejected due to photometric contamination by nearby targets. Black crosses refer to 1325 targets rejected from the fgPIC (blue F-CAM), red dots indicate a  total of 4271 targets rejected from the fgPIC (red F-CAM) due to photometric contamination from nearby Gaia\,DR3 sources. There are eight black crosses in total that do not fall exactly on a red dot (see Sect.~\ref{sec:PhotContfgPIC}).}
\label{fig:fgPICNCAM_contamination}
\end{figure}

Figure~\ref{fig:fgPICNCAM_contamination} shows the 1325 targets that were rejected from the fgPIC blue F-CAM sample and 4271 targets rejected from the fgPIC red F-CAM sample due to photometric contamination from nearby Gaia\,DR3 sources. What is readily apparent, is an increased density of rejected targets from Galactic latitudes as low as $-77^\circ.77$ towards the galactic plane near a Galactic latitude of $0^\circ$, where contamination naturally becomes an issue.

With a total count of 4279 rejected targets, the number of unique targets rejected entirely from the fgPIC due to photometric contamination is just slightly larger than the number of 4271 targets rejected from the fgPIC red F-CAM sample. This is because there are only eight targets that were rejected from the fgPIC blue F-CAM sample and that are not also rejected from the fgPIC red F-CAM sample.

\subsection{Stellar variability}

No fgPIC target is rejected because of its photometric activity. Instead, we use the phot\_variable\_flag from Gaia\,DR3 to record the photometric activity in the variable fgPICcPICvariabilityFlag (see Sect.~\ref{sec:variability}).

\section{The fgPIC}

\renewcommand{\arraystretch}{1.2}
\begin{table}[]
\caption{Definition of the bitmask values of the fgPICsourceFlag}
\begin{tabular}{p{1.58cm}|p{1.05cm}|p{3.15cm}}
{\bf Bitmask} & {\bf Integer} & {\bf fgPIC sample}  \\
\hline
0000000000 & 0 & reset all bits \\
0000000001 & 1 & star for the blue F-CAM \\
0000000010 & 2 & star for the red F-CAM \\
\end{tabular}
\label{tab:fgPICsourceFlag}
\end{table}

The final output of the selection process is the first fine-guidance PLATO Input Catalog for the LOPS2 field, version LOPS2fgPICtarget2.2.0.3. It contains a total of 2640 unique targets. This version was merged into the PIC2.2.0.1, which is published as part of PLATO's first Call for Proposals. The cPIC contains the relevant stellar identifiers, astrometric parameters, photometric information, and flags describing the properties of the selected targets. The fgPIC table is created according to the data model of the fgPIC \citep{Marrese2026}, following mostly the data model of the PIC.

As an Intermediate Data Product, the fgPIC is not published as a stand-alone file. Nevertheless, all the information about the fgPIC targets is contained in the PIC and encoded in the fgPICsourceFlag. In particular, the fgPICsourceFlag encodes the fgPIC membership and observability with the blue and red F-CAMs. The fgPICsourceFlag is stored as an integer value with the corresponding binary bitmask values shown in Table~\ref{tab:fgPICsourceFlag}.

A screening of the fgPIC reveals that 466 fgPIC targets were selected for only the blue F-CAM (fgPICsourceFlag = 1), 839 fgPIC targets were selected for only the red F-CAM (fgPICsourceFlag = 2), and 1335 additional fgPIC targets were selected for both the blue and the red F-CAMs (fgPICsourceFlag = 3).

\section{Summary}

This paper reviews the selection criteria for the cPIC and the selection criteria for the fgPIC for the first observation field of the PLATO Mission, referred to as LOPS2. Both of these catalogs were merged together with the tPIC and the scvPIC into the PIC2.2.0.1 \citep{Marrese2026}. This PIC version is published to the world as part of ESA's first call for proposals from guest observers to distribute the 8\,\% of PLATO open time offered to the scientific community.

The specific cPIC version illustrated in this paper, and merged into the PIC2.2.0.1, is the LOPS2cPICtarget2.2.0.3. It contains targets in a total of five reference stellar samples (R1, R2, R3, R4, and R5) that will be used for various technical calibration steps of the mission. The purpose of each sample is described in Table~\ref{tab:ReferenceStars}, where some targets are in multiple reference star samples. In particular, we identified nine stars that are in all cPIC samples (see Sect.~\ref{sec:output_cPIC}). In total, the cPIC has 71\,671 unique stars, which is an outcome of the cPIC selection criteria addressed in Sect.~\ref{sec:cPIC} and summarized in Table~\ref{tab:cPIC}.

For the fgPIC version LOPS2fgPICtarget2.2.0.3 that was merged into the PIC2.2.0.1, we selected a total of 2640 unique targets, 1801 of which are eligible for observations with the blue F-CAM and 2174 of which passed the selection criteria for the red F-CAM. The summary of the fgPIC selection criteria is given in Sect.~\ref{sec:selection_fgPIC} and details about the selection steps are given in Table~\ref{tab:fgPIC}. About 30 of these stars will be used by PLATO's FGS to determine the spacecraft's orientation in space and to guarantee the required exquisite pointing performance of the spacecraft.

Future versions of the cPIC and of the fgPIC, for example for the second PLATO field, may benefit from our experience with the real PLATO data. In-flight measurements of PLATO's instrument response and the predicted stellar magnitudes, the factual signal-to-noise performance of the mission, and analyses of the target flux extraction using double-aperture photometry may help to adapt the cPIC and fgPIC selection criteria to the mission's scientific goals. For example, the stellar pollution ratio may be a valuable metric to quantify the impact of photometric contaminants \citep{2026A&A...707A...2G}. Moreover, large-scale analyses of stellar variability using light curves from the TESS mission \citep{Kliapets2026} or using additional information about photometric variability from {\it Gaia} \citep{2023A&A...677A.137M} could help to improve the selection of photometrically stable cPIC and fgPIC targets.

\bmhead{Acknowledgments}

% Proof-reading
The authors thank Conny Aerts and Pierre Royer for their valuable comments on a draft version of this manuscript.
% PLATO standard acknowledgement
This work presents results from the European Space Agency (ESA) space mission PLATO. The PLATO payload is jointly developed by ESA and the PLATO Mission Consortium (PMC). The PMC furthermore contributes to the mission science ground segment processing PLATO data. Funding for the PMC is provided by national institutions, in particular the institutions participating in the PLATO Multi-Lateral Agreement (MLA) and represented in the PMC board (Austria, Belgium, Brazil, Czech Republic, Denmark, France, Germany, Italy, Netherlands, Norway, Portugal, Spain, Sweden, Switzerland, United Kingdom). Members of the PMC can be found at the consortium website: https://platomission.com/. The ESA PLATO mission website is https://www.cosmos.esa.int/plato. We thank the teams working for PLATO for all their efforts.
% \textit{Gaia} Mission acknowleddgement
This work has made use of data from the European Space Agency (ESA) mission
\textit{Gaia} (\url{https://www.cosmos.esa.int/gaia}), processed by the Gaia Data Processing and Analysis Consortium (DPAC,
\url{https://www.cosmos.esa.int/web/gaia/dpac/consortium}). Funding for the DPAC has been provided by national institutions, in particular the institutions participating in the Gaia Multilateral Agreement.
% Authors' funding acknowledgements.
% *?*
R.~Heller, C.~Jiang, M.~Sch{\"a}fer, M.~Ammler-von Eiff, C.~Damiani and C.~Rauterberg acknowledge support from the German Aerospace Agency (Deutsches Zentrum f{\"u}r Luft- und Raumfahrt) under PLATO Data Center grants 50OO1501 and 50OP1902.
V.~Granata, G.~Piotto, and M.~Montalto acknowledge support from PLATO ASI-INAF agreements n. 2022-28-HH.0
This work was shaped in important ways by the insight and commitment of Patrick Gaulme. He passed away in July 2025 at a far too young age and is deeply missed.


%\bibliographystyle{bst/sn-basic}
\bibliography{bibliography}

@ARTICLE{2025ExA....59...26R,
       author = {{Rauer}, Heike and {Aerts}, Conny and {Cabrera}, Juan and {Deleuil}, Magali and {Erikson}, Anders and {Gizon}, Laurent and {Goupil}, Mariejo and {Heras}, Ana and {Walloschek}, Thomas and {Lorenzo-Alvarez}, Jose and {Marliani}, Filippo and {Martin-Garcia}, C{\'e}sar and {Mas-Hesse}, J. Miguel and {O'Rourke}, Laurence and {Osborn}, Hugh and {Pagano}, Isabella and {Piotto}, Giampaolo and {Pollacco}, Don and {Ragazzoni}, Roberto and {Ramsay}, Gavin and {Udry}, St{\'e}phane and {Appourchaux}, Thierry and {Benz}, Willy and {Brandeker}, Alexis and {G{\"u}del}, Manuel and {Janot-Pacheco}, Eduardo and {Kabath}, Petr and {Kjeldsen}, Hans and {Min}, Michiel and {Santos}, Nuno and {Smith}, Alan and {Suarez}, Juan-Carlos and {Werner}, Stephanie C. and {Aboudan}, Alessio and {Abreu}, Manuel and {Acu{\~n}a}, Lorena and {Adams}, Moritz and {Adibekyan}, Vardan and {Affer}, Laura and {Agneray}, Fran{\c{c}}ois and {Agnor}, Craig and {Aguirre B{\o}rsen-Koch}, Victor and {Ahmed}, Saad and {Aigrain}, Suzanne and {Al-Bahlawan}, Ashraf and {Alcacera Gil}, Ma de los Angeles and {Alei}, Eleonora and {Alencar}, Silvia and {Alexander}, Richard and {Alfonso-Garz{\'o}n}, Julia and {Alibert}, Yann and {Allende Prieto}, Carlos and {Almeida}, Leonardo and {Alonso Sobrino}, Roi and {Altavilla}, Giuseppe and {Althaus}, Christian and {Alvarez Trujillo}, Luis Alonso and {Amarsi}, Anish and {Ammler-von Eiff}, Matthias and {Am{\^o}res}, Eduardo and {Andrade}, Laerte and {Antoniadis-Karnavas}, Alexandros and {Ant{\'o}nio}, Carlos and {Aparicio del Moral}, Beatriz and {Appolloni}, Matteo and {Arena}, Claudio and {Armstrong}, David and {Aroca Aliaga}, Jose and {Asplund}, Martin and {Audenaert}, Jeroen and {Auricchio}, Natalia and {Avelino}, Pedro and {Baeke}, Ann and {Bailli{\'e}}, Kevin and {Balado}, Ana and {Ballber Balaguer{\'o}}, Pau and {Balestra}, Andrea and {Ball}, Warrick and {Ballans}, Herve and {Ballot}, Jerome and {Barban}, Caroline and {Barbary}, Ga{\"e}le and {Barbieri}, Mauro and {Barcel{\'o} Forteza}, Sebasti{\`a} and {Barker}, Adrian and {Barklem}, Paul and {Barnes}, Sydney and {Barrado Navascues}, David and {Barragan}, Oscar and {Baruteau}, Cl{\'e}ment and {Basu}, Sarbani and {Baudin}, Frederic and {Baumeister}, Philipp and {Bayliss}, Daniel and {Bazot}, Michael and {Beck}, Paul G. and {Belkacem}, Kevin and {Bellinger}, Earl and {Benatti}, Serena and {Benomar}, Othman and {B{\'e}rard}, Diane and {Bergemann}, Maria and {Bergomi}, Maria and {Bernardo}, Pierre and {Biazzo}, Katia and {Bignamini}, Andrea and {Bigot}, Lionel and {Billot}, Nicolas and {Binet}, Martin and {Biondi}, David and {Biondi}, Federico and {Birch}, Aaron C. and {Bitsch}, Bertram and {Bluhm Ceballos}, Paz Victoria and {B{\'o}di}, Attila and {Bogn{\'a}r}, Zs{\'o}fia and {Boisse}, Isabelle and {Bolmont}, Emeline and {Bonanno}, Alfio and {Bonavita}, Mariangela and {Bonfanti}, Andrea and {Bonfils}, Xavier and {Bonito}, Rosaria and {Bonomo}, Aldo Stefano and {B{\"o}rner}, Anko and {Boro Saikia}, Sudeshna and {Borreguero Mart{\'\i}n}, Elisa and {Borsa}, Francesco and {Borsato}, Luca and {Bossini}, Diego and {Bouchy}, Francois and {Bou{\'e}}, Gwena{\"e}l and {Boufleur}, Rodrigo and {Boumier}, Patrick and {Bourrier}, Vincent and {Bowman}, Dominic M. and {Bozzo}, Enrico and {Bradley}, Louisa and {Bray}, John and {Bressan}, Alessandro and {Breton}, Sylvain and {Brienza}, Daniele and {Brito}, Ana and {Brogi}, Matteo and {Brown}, Beverly and {Brown}, David J.~A. and {Brun}, Allan Sacha and {Bruno}, Giovanni and {Bruns}, Michael and {Buchhave}, Lars A. and {Bugnet}, Lisa and {Buldgen}, Ga{\"e}l and {Burgess}, Patrick and {Busatta}, Andrea and {Busso}, Giorgia and {Buzasi}, Derek and {Caballero}, Jos{\'e} A. and {Cabral}, Alexandre and {Cabrero Gomez}, Juan-Francisco and {Calderone}, Flavia and {Cameron}, Robert and {Cameron}, Andrew and {Campante}, Tiago and {Campos Gestal}, N{\'e}stor and {Canto Martins}, Bruno Leonardo and {Cara}, Christophe and {Carone}, Ludmila and {Carrasco}, Josep Manel and {Casagrande}, Luca and {Casewell}, Sarah L. and {Cassisi}, Santi and {Castellani}, Marco and {Castro}, Matthieu and {Catala}, Claude and {Catal{\'a}n Fern{\'a}ndez}, Irene and {Catelan}, M{\'a}rcio and {Cegla}, Heather and {Cerruti}, Chiara and {Cessa}, Virginie and {Chadid}, Merieme and {Chaplin}, William and {Charpinet}, Stephane and {Chiappini}, Cristina and {Chiarucci}, Simone and {Chiavassa}, Andrea and {Chinellato}, Simonetta and {Chirulli}, Giovanni and {Christensen-Dalsgaard}, J{\o}rgen and {Church}, Ross and {Claret}, Antonio and {Clarke}, Cathie and {Claudi}, Riccardo and {Clermont}, Lionel and {Coelho}, Hugo and {Coelho}, Joao and {Cogato}, Fabrizio and {Colom{\'e}}, Josep and {Condamin}, Mathieu and {Conde Garc{\'\i}a}, Fernando and {Conseil}, Simon},
        title = "{The PLATO mission}",
      journal = {Experimental Astronomy},
         year = 2025,
        month = jun,
       volume = {59},
       number = {3},
          eid = {26},
        pages = {26},
          doi = {10.1007/s10686-025-09985-9},
archivePrefix = {arXiv},
       eprint = {2406.05447},
 primaryClass = {astro-ph.IM},
       adsurl = {https://ui.adsabs.harvard.edu/abs/2025ExA....59...26R}
}

@ARTICLE{Zwintz2026,
       author = {{Zwintz}, Konstanze and {Aerts}, Conny and {Tkachenko}, Andrew and {Cabrera}, Juan and {Creevey}, Orlagh and {Heller}, Ren{\'e} and {Jannsen}, Nicholas and {Jiang}, Chen and {Kochukhov}, Oleg and {Lanza}, Antonino Francesco and {Maxted}, Pierre F. L. and {Messina}, Sergio and {Miglio}, Andrea and {Morel}, Thierry},
        title = "{The PLATO Science Calibration and Validation Plan: Targets for the First Long-pointing Field}",
      journal = {Experimental Astronomy (in prep.)},
         year = 2026
}

@ARTICLE{Montalto2026,
       author = {{Montalto}, Marco and {Piotto} Giampaolo and {Marrese} Paola Maria and {Prisinzano} L. and {Marinoni} Silvia and {Granata} Valentina and {Cabrera} Juan and {Nascimbeni} Valerio and {Desidera} A. and {Vardan} S. and {Ortolani} S. and {Alei} E. and {Aerts} C. and {Altavilla} G. and {Belkacem} K. and {Benatti} S. and {Börner} A. and {Deleuil} M. and {Gizon} Laurent and {Goupil} M.J. and {Günther} M. and {Heras} A.M. and {Magrin} D. and {Malavolta} L and {Mas-Hesse} J.M. and {Pagano} Isabella and {Paproth} Carsten and {Pollaco} Don and {Ragazzoni} R. and {Ramsay} Gavin and {Rauer} Heike and {Udry} Stéphane},
        title = "{The PLATO Input Catalogue of targets (tPIC)}",
      journal = {Experimental Astronomy (in prep.)},
         year = 2026
}

@ARTICLE{2024A&A...692A.150B,
       author = {{Boettner}, C. and {Viswanathan}, A. and {Dayal}, P.},
        title = "{Exoplanets across galactic stellar populations with PLATO: Estimating exoplanet yields around FGK stars for the thin disk, thick disk, and stellar halo}",
      journal = {\aap},
         year = 2024,
        month = dec,
       volume = {692},
          eid = {A150},
        pages = {A150},
          doi = {10.1051/0004-6361/202451537},
archivePrefix = {arXiv},
       eprint = {2407.15917},
 primaryClass = {astro-ph.EP},
       adsurl = {https://ui.adsabs.harvard.edu/abs/2024A&A...692A.150B}
}

@ARTICLE{Kliapets2026,
       author = {{Kliapets}, Mykyta and {Huyse}, Pablo and {Audenaert}, Jeroen~ and {et al.} },
        title = "{Variability classification of TESS targets in LOPS2, the first long-term pointing field of PLATO. Version 1 of the public variability catalogue}",
      journal = {Astronomy \& Astrophysics (in prep.)},
         year = 2026
}

@techreport{fgPIC2026,
  author      = {{Heller}, Ren{\'e} and {Jiang}, Chen},
  title       = "Technical Verification Document for the fg{PIC} for {PIC} v2.2 in {LOPS2}",
  institution = "Max Planck Institute for Solar System Research",
  year        = "2026",
  type        = "Technical Note",
  number      = "i1.0",
  month       = "Feb.",
  note        = "167 pp. Available with restrictions upon request to the corresponding author."
}

@techreport{cPIC2026,
  author      = {{Heller}, Ren{\'e} and {Jiang}, Chen},
  title       = "Technical Verification Document for the c{PIC} for {PIC} v2.2 in {LOPS2}",
  institution = "Max Planck Institute for Solar System Research",
  year        = "2026",
  type        = "Technical Note",
  number      = "i1.0",
  month       = "Feb.",
  note        = "24 pp. Available with restrictions upon request to the corresponding author."
}

@techreport{scvPIC2026,
  author      = {{Heller}, Ren{\'e} and {Jiang}, Chen},
  title       = "Technical Verification Document for the scv{PIC} for {PIC} v2.2 in {LOPS2}",
  institution = "Max Planck Institute for Solar System Research",
  year        = "2026",
  type        = "Technical Note",
  number      = "i1.0",
  month       = "Mar.",
  note        = "18 pp. Available with restrictions upon request to the corresponding author."
}

@ARTICLE{2010A&A...523A..48J,
       author = {{Jordi}, C. and {Gebran}, M. and {Carrasco}, J.~M. and {de Bruijne}, J. and {Voss}, H. and {Fabricius}, C. and {Knude}, J. and {Vallenari}, A. and {Kohley}, R. and {Mora}, A.},
        title = "{Gaia broad band photometry}",
      journal = {\aap},
         year = 2010,
        month = nov,
       volume = {523},
          eid = {A48},
        pages = {A48},
          doi = {10.1051/0004-6361/201015441},
archivePrefix = {arXiv},
       eprint = {1008.0815},
 primaryClass = {astro-ph.IM},
       adsurl = {https://ui.adsabs.harvard.edu/abs/2010A&A...523A..48J}
}

@ARTICLE{2026arXiv260312750L,
       author = {{Lund}, Mikkel N. and {Ballot}, J{\'e}r{\^o}me and {Chaplin}, William J.},
        title = "{Bolometric corrections of stellar oscillation mode amplitudes as observed by the PLATO mission. I. Planck-spectrum estimates}",
      journal = {arXiv e-prints},
         year = 2026,
        month = mar,
          eid = {arXiv:2603.12750},
        pages = {arXiv:2603.12750},
          doi = {10.48550/arXiv.2603.12750},
archivePrefix = {arXiv},
       eprint = {2603.12750},
 primaryClass = {astro-ph.SR},
       adsurl = {https://ui.adsabs.harvard.edu/abs/2026arXiv260312750L}
}

@techreport{Montalto2024,
  author      = {{Montalto}, Marco},
  title       = "PIC2PARAM v3.4 user manual",
  institution = "Padova University",
  year        = "2024",
  type        = "Technical Note",
  number      = "i1.0",
  month       = "Apr.",
  note        = "{P}LATO-UPD-SCI-TN-0023, 16 pp. Available with restrictions upon request to the corresponding author."
}

@ARTICLE{2026arXiv260305586B,
       author = {{Boyle}, Andrew W. and {Bouma}, Luke G. and {Mann}, Andrew W.},
        title = "{The TESS All-Sky Rotation Survey: Periods for 944,056 Stars Within 500 pc}",
      journal = {arXiv e-prints},
         year = 2026,
        month = mar,
          eid = {arXiv:2603.05586},
        pages = {arXiv:2603.05586},
archivePrefix = {arXiv},
       eprint = {2603.05586},
 primaryClass = {astro-ph.SR},
       adsurl = {https://ui.adsabs.harvard.edu/abs/2026arXiv260305586B}
}

@ARTICLE{2024ExA....58....1B,
       author = {{B{\"o}rner}, Anko and {Paproth}, Carsten and {Cabrera}, Juan and {Pertenais}, Martin and {Rauer}, Heike and {Mas-Hesse}, J. Miguel and {Pagano}, Isabella and {Alvarez}, Jose Lorenzo and {Erikson}, Anders and {Grie{\ss}bach}, Denis and {Levillain}, Yves and {Magrin}, Demetrio and {Mogulsky}, Valery and {Niemi}, Sami-Matias and {Prod'homme}, Thibaut and {Regibo}, Sara and {De Ridder}, Joris and {Rockstein}, Steve and {Samadi}, Reza and {Serrano-Velarde}, Dimitri and {Smith}, Alan and {Verhoeve}, Peter and {Walton}, Dave},
        title = "{PLATO's signal and noise budget}",
      journal = {Experimental Astronomy},
         year = 2024,
        month = aug,
       volume = {58},
       number = {1},
          eid = {1},
        pages = {1},
          doi = {10.1007/s10686-024-09948-6},
archivePrefix = {arXiv},
       eprint = {2406.11556},
 primaryClass = {astro-ph.EP},
       adsurl = {https://ui.adsabs.harvard.edu/abs/2024ExA....58....1B}
}

@ARTICLE{2019LRSP...16....4G,
       author = {{Garc{\'\i}a}, Rafael A. and {Ballot}, J{\'e}r{\^o}me},
        title = "{Asteroseismology of solar-type stars}",
      journal = {Living Reviews in Solar Physics},
         year = 2019,
        month = dec,
       volume = {16},
       number = {1},
          eid = {4},
        pages = {4},
          doi = {10.1007/s41116-019-0020-1},
archivePrefix = {arXiv},
       eprint = {1906.12262},
 primaryClass = {astro-ph.SR},
       adsurl = {https://ui.adsabs.harvard.edu/abs/2019LRSP...16....4G}
}

@ARTICLE{2023A&A...677A.137M,
       author = {{Ma{\'\i}z Apell{\'a}niz}, J. and {Holgado}, G. and {Pantaleoni Gonz{\'a}lez}, M. and {Caballero}, J.~A.},
        title = "{Stellar variability in Gaia DR3. I. Three-band photometric dispersions for 145 million sources}",
      journal = {\aap},
         year = 2023,
        month = sep,
       volume = {677},
          eid = {A137},
        pages = {A137},
          doi = {10.1051/0004-6361/202346759},
archivePrefix = {arXiv},
       eprint = {2304.14249},
 primaryClass = {astro-ph.SR},
       adsurl = {https://ui.adsabs.harvard.edu/abs/2023A&A...677A.137M}
}

@ARTICLE{2023A&A...677A.133M,
       author = {{Matuszewski}, F. and {Nettelmann}, N. and {Cabrera}, J. and {B{\"o}rner}, A. and {Rauer}, H.},
        title = "{Estimating the number of planets that PLATO can detect}",
      journal = {\aap},
         year = 2023,
        month = sep,
       volume = {677},
          eid = {A133},
        pages = {A133},
          doi = {10.1051/0004-6361/202245287},
archivePrefix = {arXiv},
       eprint = {2307.12163},
 primaryClass = {astro-ph.EP},
       adsurl = {https://ui.adsabs.harvard.edu/abs/2023A&A...677A.133M}
}

@ARTICLE{Marrese2026,
       author = {{Marrese}, Paola~Maria and {Marrinoni}, Silvia~ and {et al.} },
        title = "{Implementation of the PLATO Input Catalog}",
      journal = {Experimental Astronomy (in prep.)},
         year = 2026
}

@ARTICLE{2016A&A...595A...1G,
       author = {{Gaia Collaboration} and {Prusti}, T. and {de Bruijne}, J.~H.~J. and {Brown}, A.~G.~A. and {Vallenari}, A. and {Babusiaux}, C. and {Bailer-Jones}, C.~A.~L. and {Bastian}, U. and {Biermann}, M. and {Evans}, D.~W. and {Eyer}, L. and {Jansen}, F. and {Jordi}, C. and {Klioner}, S.~A. and {Lammers}, U. and {Lindegren}, L. and {Luri}, X. and {Mignard}, F. and {Milligan}, D.~J. and {Panem}, C. and {Poinsignon}, V. and {Pourbaix}, D. and {Randich}, S. and {Sarri}, G. and {Sartoretti}, P. and {Siddiqui}, H.~I. and {Soubiran}, C. and {Valette}, V. and {van Leeuwen}, F. and {Walton}, N.~A. and {Aerts}, C. and {Arenou}, F. and {Cropper}, M. and {Drimmel}, R. and {H{\o}g}, E. and {Katz}, D. and {Lattanzi}, M.~G. and {O'Mullane}, W. and {Grebel}, E.~K. and {Holland}, A.~D. and {Huc}, C. and {Passot}, X. and {Bramante}, L. and {Cacciari}, C. and {Casta{\~n}eda}, J. and {Chaoul}, L. and {Cheek}, N. and {De Angeli}, F. and {Fabricius}, C. and {Guerra}, R. and {Hern{\'a}ndez}, J. and {Jean-Antoine-Piccolo}, A. and {Masana}, E. and {Messineo}, R. and {Mowlavi}, N. and {Nienartowicz}, K. and {Ord{\'o}{\~n}ez-Blanco}, D. and {Panuzzo}, P. and {Portell}, J. and {Richards}, P.~J. and {Riello}, M. and {Seabroke}, G.~M. and {Tanga}, P. and {Th{\'e}venin}, F. and {Torra}, J. and {Els}, S.~G. and {Gracia-Abril}, G. and {Comoretto}, G. and {Garcia-Reinaldos}, M. and {Lock}, T. and {Mercier}, E. and {Altmann}, M. and {Andrae}, R. and {Astraatmadja}, T.~L. and {Bellas-Velidis}, I. and {Benson}, K. and {Berthier}, J. and {Blomme}, R. and {Busso}, G. and {Carry}, B. and {Cellino}, A. and {Clementini}, G. and {Cowell}, S. and {Creevey}, O. and {Cuypers}, J. and {Davidson}, M. and {De Ridder}, J. and {de Torres}, A. and {Delchambre}, L. and {Dell'Oro}, A. and {Ducourant}, C. and {Fr{\'e}mat}, Y. and {Garc{\'\i}a-Torres}, M. and {Gosset}, E. and {Halbwachs}, J.-L. and {Hambly}, N.~C. and {Harrison}, D.~L. and {Hauser}, M. and {Hestroffer}, D. and {Hodgkin}, S.~T. and {Huckle}, H.~E. and {Hutton}, A. and {Jasniewicz}, G. and {Jordan}, S. and {Kontizas}, M. and {Korn}, A.~J. and {Lanzafame}, A.~C. and {Manteiga}, M. and {Moitinho}, A. and {Muinonen}, K. and {Osinde}, J. and {Pancino}, E. and {Pauwels}, T. and {Petit}, J.-M. and {Recio-Blanco}, A. and {Robin}, A.~C. and {Sarro}, L.~M. and {Siopis}, C. and {Smith}, M. and {Smith}, K.~W. and {Sozzetti}, A. and {Thuillot}, W. and {van Reeven}, W. and {Viala}, Y. and {Abbas}, U. and {Abreu Aramburu}, A. and {Accart}, S. and {Aguado}, J.~J. and {Allan}, P.~M. and {Allasia}, W. and {Altavilla}, G. and {{\'A}lvarez}, M.~A. and {Alves}, J. and {Anderson}, R.~I. and {Andrei}, A.~H. and {Anglada Varela}, E. and {Antiche}, E. and {Antoja}, T. and {Ant{\'o}n}, S. and {Arcay}, B. and {Atzei}, A. and {Ayache}, L. and {Bach}, N. and {Baker}, S.~G. and {Balaguer-N{\'u}{\~n}ez}, L. and {Barache}, C. and {Barata}, C. and {Barbier}, A. and {Barblan}, F. and {Baroni}, M. and {Barrado y Navascu{\'e}s}, D. and {Barros}, M. and {Barstow}, M.~A. and {Becciani}, U. and {Bellazzini}, M. and {Bellei}, G. and {Bello Garc{\'\i}a}, A. and {Belokurov}, V. and {Bendjoya}, P. and {Berihuete}, A. and {Bianchi}, L. and {Bienaym{\'e}}, O. and {Billebaud}, F. and {Blagorodnova}, N. and {Blanco-Cuaresma}, S. and {Boch}, T. and {Bombrun}, A. and {Borrachero}, R. and {Bouquillon}, S. and {Bourda}, G. and {Bouy}, H. and {Bragaglia}, A. and {Breddels}, M.~A. and {Brouillet}, N. and {Br{\"u}semeister}, T. and {Bucciarelli}, B. and {Budnik}, F. and {Burgess}, P. and {Burgon}, R. and {Burlacu}, A. and {Busonero}, D. and {Buzzi}, R. and {Caffau}, E. and {Cambras}, J. and {Campbell}, H. and {Cancelliere}, R. and {Cantat-Gaudin}, T. and {Carlucci}, T. and {Carrasco}, J.~M. and {Castellani}, M. and {Charlot}, P. and {Charnas}, J. and {Charvet}, P. and {Chassat}, F. and {Chiavassa}, A. and {Clotet}, M. and {Cocozza}, G. and {Collins}, R.~S. and {Collins}, P. and {Costigan}, G.},
        title = "{The Gaia mission}",
      journal = {\aap},
         year = 2016,
        month = nov,
       volume = {595},
          eid = {A1},
        pages = {A1},
          doi = {10.1051/0004-6361/201629272},
archivePrefix = {arXiv},
       eprint = {1609.04153},
 primaryClass = {astro-ph.IM},
       adsurl = {https://ui.adsabs.harvard.edu/abs/2016A&A...595A...1G}
}

@ARTICLE{2023A&A...674A...1G,
       author = {{Gaia Collaboration} and {Vallenari}, A. and {Brown}, A.~G.~A. and {Prusti}, T. and {de Bruijne}, J.~H.~J. and {Arenou}, F. and {Babusiaux}, C. and {Biermann}, M. and {Creevey}, O.~L. and {Ducourant}, C. and {Evans}, D.~W. and {Eyer}, L. and {Guerra}, R. and {Hutton}, A. and {Jordi}, C. and {Klioner}, S.~A. and {Lammers}, U.~L. and {Lindegren}, L. and {Luri}, X. and {Mignard}, F. and {Panem}, C. and {Pourbaix}, D. and {Randich}, S. and {Sartoretti}, P. and {Soubiran}, C. and {Tanga}, P. and {Walton}, N.~A. and {Bailer-Jones}, C.~A.~L. and {Bastian}, U. and {Drimmel}, R. and {Jansen}, F. and {Katz}, D. and {Lattanzi}, M.~G. and {van Leeuwen}, F. and {Bakker}, J. and {Cacciari}, C. and {Casta{\~n}eda}, J. and {De Angeli}, F. and {Fabricius}, C. and {Fouesneau}, M. and {Fr{\'e}mat}, Y. and {Galluccio}, L. and {Guerrier}, A. and {Heiter}, U. and {Masana}, E. and {Messineo}, R. and {Mowlavi}, N. and {Nicolas}, C. and {Nienartowicz}, K. and {Pailler}, F. and {Panuzzo}, P. and {Riclet}, F. and {Roux}, W. and {Seabroke}, G.~M. and {Sordo}, R. and {Th{\'e}venin}, F. and {Gracia-Abril}, G. and {Portell}, J. and {Teyssier}, D. and {Altmann}, M. and {Andrae}, R. and {Audard}, M. and {Bellas-Velidis}, I. and {Benson}, K. and {Berthier}, J. and {Blomme}, R. and {Burgess}, P.~W. and {Busonero}, D. and {Busso}, G. and {C{\'a}novas}, H. and {Carry}, B. and {Cellino}, A. and {Cheek}, N. and {Clementini}, G. and {Damerdji}, Y. and {Davidson}, M. and {de Teodoro}, P. and {Nu{\~n}ez Campos}, M. and {Delchambre}, L. and {Dell'Oro}, A. and {Esquej}, P. and {Fern{\'a}ndez-Hern{\'a}ndez}, J. and {Fraile}, E. and {Garabato}, D. and {Garc{\'\i}a-Lario}, P. and {Gosset}, E. and {Haigron}, R. and {Halbwachs}, J.-L. and {Hambly}, N.~C. and {Harrison}, D.~L. and {Hern{\'a}ndez}, J. and {Hestroffer}, D. and {Hodgkin}, S.~T. and {Holl}, B. and {Jan{\ss}en}, K. and {Jevardat de Fombelle}, G. and {Jordan}, S. and {Krone-Martins}, A. and {Lanzafame}, A.~C. and {L{\"o}ffler}, W. and {Marchal}, O. and {Marrese}, P.~M. and {Moitinho}, A. and {Muinonen}, K. and {Osborne}, P. and {Pancino}, E. and {Pauwels}, T. and {Recio-Blanco}, A. and {Reyl{\'e}}, C. and {Riello}, M. and {Rimoldini}, L. and {Roegiers}, T. and {Rybizki}, J. and {Sarro}, L.~M. and {Siopis}, C. and {Smith}, M. and {Sozzetti}, A. and {Utrilla}, E. and {van Leeuwen}, M. and {Abbas}, U. and {{\'A}brah{\'a}m}, P. and {Abreu Aramburu}, A. and {Aerts}, C. and {Aguado}, J.~J. and {Ajaj}, M. and {Aldea-Montero}, F. and {Altavilla}, G. and {{\'A}lvarez}, M.~A. and {Alves}, J. and {Anders}, F. and {Anderson}, R.~I. and {Anglada Varela}, E. and {Antoja}, T. and {Baines}, D. and {Baker}, S.~G. and {Balaguer-N{\'u}{\~n}ez}, L. and {Balbinot}, E. and {Balog}, Z. and {Barache}, C. and {Barbato}, D. and {Barros}, M. and {Barstow}, M.~A. and {Bartolom{\'e}}, S. and {Bassilana}, J.-L. and {Bauchet}, N. and {Becciani}, U. and {Bellazzini}, M. and {Berihuete}, A. and {Bernet}, M. and {Bertone}, S. and {Bianchi}, L. and {Binnenfeld}, A. and {Blanco-Cuaresma}, S. and {Blazere}, A. and {Boch}, T. and {Bombrun}, A. and {Bossini}, D. and {Bouquillon}, S. and {Bragaglia}, A. and {Bramante}, L. and {Breedt}, E. and {Bressan}, A. and {Brouillet}, N. and {Brugaletta}, E. and {Bucciarelli}, B. and {Burlacu}, A. and {Butkevich}, A.~G. and {Buzzi}, R. and {Caffau}, E. and {Cancelliere}, R. and {Cantat-Gaudin}, T. and {Carballo}, R. and {Carlucci}, T. and {Carnerero}, M.~I. and {Carrasco}, J.~M. and {Casamiquela}, L. and {Castellani}, M. and {Castro-Ginard}, A. and {Chaoul}, L. and {Charlot}, P. and {Chemin}, L. and {Chiaramida}, V. and {Chiavassa}, A. and {Chornay}, N. and {Comoretto}, G. and {Contursi}, G. and {Cooper}, W.~J. and {Cornez}, T. and {Cowell}, S. and {Crifo}, F. and {Cropper}, M. and {Crosta}, M. and {Crowley}, C. and {Dafonte}, C. and {Dapergolas}, A. and {David}, M. and {David}, P. and {de Laverny}, P. and {De Luise}, F. and {De March}, R.},
        title = "{Gaia Data Release 3. Summary of the content and survey properties}",
      journal = {\aap},
         year = 2023,
        month = jun,
       volume = {674},
          eid = {A1},
        pages = {A1},
          doi = {10.1051/0004-6361/202243940},
archivePrefix = {arXiv},
       eprint = {2208.00211},
 primaryClass = {astro-ph.GA},
       adsurl = {https://ui.adsabs.harvard.edu/abs/2023A&A...674A...1G}
}

@INPROCEEDINGS{2022SPIE12180E..4MP,
       author = {{Pertenais}, Martin and {Ammler-von Eiff}, Matthias and {Burresi}, Matteo and {Cabrera}, Juan and {Farinato}, Jacopo and {Gorius}, Nicolas and {Huygen}, Rik and {Magrin}, Demetrio and {Martin Garcia}, Cesar and {Munari}, Matteo and {Regibo}, Sara and {Royer}, Pierre and {Vandenbussche}, Bart and {van Kempen}, Tim A.},
        title = "{PLATO camera ghosts: simulations and measurements on the engineering model (EM)}",
    booktitle = {Space Telescopes and Instrumentation 2022: Optical, Infrared, and Millimeter Wave},
         year = 2022,
       editor = {{Coyle}, Laura E. and {Matsuura}, Shuji and {Perrin}, Marshall D.},
       series = {Society of Photo-Optical Instrumentation Engineers (SPIE) Conference Series},
       volume = {12180},
        month = aug,
          eid = {121804M},
        pages = {121804M},
          doi = {10.1117/12.2629189},
       adsurl = {https://ui.adsabs.harvard.edu/abs/2022SPIE12180E..4MP}
}

@INPROCEEDINGS{2021SPIE11852E..4YP,
       author = {{Pertenais}, Martin and {Cabrera}, Juan and {Paproth}, Carsten and {Boerner}, Anko and {Grie{\ss}bach}, Denis and {Mogulsky}, Valery and {Rauer}, Heike},
        title = "{The unique field-of-view and focusing budgets of PLATO}",
    booktitle = {International Conference on Space Optics {\textemdash} ICSO 2020},
         year = 2021,
       editor = {{Cugny}, Bruno and {Sodnik}, Zoran and {Karafolas}, Nikos},
       series = {Society of Photo-Optical Instrumentation Engineers (SPIE) Conference Series},
       volume = {11852},
        month = jun,
          eid = {118524Y},
        pages = {118524Y},
          doi = {10.1117/12.2599820},
       adsurl = {https://ui.adsabs.harvard.edu/abs/2021SPIE11852E..4YP}
}

@ARTICLE{2025arXiv251022092M,
       author = {{Mishra}, Shaunak and {Samadi}, Reza and {B{\'e}rard}, Diane},
        title = "{Impact of Charge Transfer Inefficiency on transit light-curves: A correction strategy for PLATO}",
      journal = {arXiv e-prints},
         year = 2025,
        month = oct,
          eid = {arXiv:2510.22092},
        pages = {arXiv:2510.22092},
          doi = {10.48550/arXiv.2510.22092},
archivePrefix = {arXiv},
       eprint = {2510.22092},
 primaryClass = {astro-ph.EP},
       adsurl = {https://ui.adsabs.harvard.edu/abs/2025arXiv251022092M}
}

@ARTICLE{2019A&A...627A..71M,
       author = {{Marchiori}, V. and {Samadi}, R. and {Fialho}, F. and {Paproth}, C. and {Santerne}, A. and {Pertenais}, M. and {B{\"o}rner}, A. and {Cabrera}, J. and {Monsky}, A. and {Kutrowski}, N.},
        title = "{In-flight photometry extraction of PLATO targets. Optimal apertures for detecting extrasolar planets}",
      journal = {\aap},
         year = 2019,
        month = jul,
       volume = {627},
          eid = {A71},
        pages = {A71},
          doi = {10.1051/0004-6361/201935269},
archivePrefix = {arXiv},
       eprint = {1906.00892},
 primaryClass = {astro-ph.IM},
       adsurl = {https://ui.adsabs.harvard.edu/abs/2019A&A...627A..71M}
}

@ARTICLE{Cabrera2026,
       author = {{Cabrera}, Juan and {Rauer}, Heike and {Samadi}, R{\'e}za and {Nascimbeni}, Valerio and {B{\"o}rner}, Anko and {Grie{\ss}bach}, Denis and {Paproth}, Carsten},
        title = "{Assessment of PLATO Science Performance}",
      journal = {Experimental Astronomy (in prep.)},
         year = 2026
}

@ARTICLE{2017ApJ...835..172L,
       author = {{Lund}, Mikkel N. and {Silva Aguirre}, V{\'\i}ctor and {Davies}, Guy R. and {Chaplin}, William J. and {Christensen-Dalsgaard}, J{\o}rgen and {Houdek}, G{\"u}nter and {White}, Timothy R. and {Bedding}, Timothy R. and {Ball}, Warrick H. and {Huber}, Daniel and {Antia}, H.~M. and {Lebreton}, Yveline and {Latham}, David W. and {Handberg}, Rasmus and {Verma}, Kuldeep and {Basu}, Sarbani and {Casagrande}, Luca and {Justesen}, Anders B. and {Kjeldsen}, Hans and {Mosumgaard}, Jakob R.},
        title = "{Standing on the Shoulders of Dwarfs: the Kepler Asteroseismic LEGACY Sample. I. Oscillation Mode Parameters}",
      journal = {\apj},
         year = 2017,
        month = feb,
       volume = {835},
       number = {2},
          eid = {172},
        pages = {172},
          doi = {10.3847/1538-4357/835/2/172},
archivePrefix = {arXiv},
       eprint = {1612.00436},
 primaryClass = {astro-ph.SR},
       adsurl = {https://ui.adsabs.harvard.edu/abs/2017ApJ...835..172L}
}

@ARTICLE{2019A&A...624A.117S,
       author = {{Samadi}, R. and {Deru}, A. and {Reese}, D. and {Marchiori}, V. and {Grolleau}, E. and {Green}, J.~J. and {Pertenais}, M. and {Lebreton}, Y. and {Deheuvels}, S. and {Mosser}, B. and {Belkacem}, K. and {B{\"o}rner}, A. and {Smith}, A.~M.~S.},
        title = "{The PLATO Solar-like Light-curve Simulator. A tool to generate realistic stellar light-curves with instrumental effects representative of the PLATO mission}",
      journal = {\aap},
         year = 2019,
        month = apr,
       volume = {624},
          eid = {A117},
        pages = {A117},
          doi = {10.1051/0004-6361/201834822},
archivePrefix = {arXiv},
       eprint = {1903.02747},
 primaryClass = {astro-ph.IM},
       adsurl = {https://ui.adsabs.harvard.edu/abs/2019A&A...624A.117S}
}

@ARTICLE{2014A&A...566A..92M,
       author = {{Marcos-Arenal}, P. and {Zima}, W. and {De Ridder}, J. and {Aerts}, C. and {Huygen}, R. and {Samadi}, R. and {Green}, J. and {Piotto}, G. and {Salmon}, S. and {Catala}, C. and {Rauer}, H.},
        title = "{The PLATO Simulator: modelling of high-precision high-cadence space-based imaging}",
      journal = {\aap},
         year = 2014,
        month = jun,
       volume = {566},
          eid = {A92},
        pages = {A92},
          doi = {10.1051/0004-6361/201323304},
archivePrefix = {arXiv},
       eprint = {1404.1886},
 primaryClass = {astro-ph.IM},
       adsurl = {https://ui.adsabs.harvard.edu/abs/2014A&A...566A..92M}
}

@ARTICLE{2024A&A...681A..18J,
       author = {{Jannsen}, N. and {De Ridder}, J. and {Seynaeve}, D. and {Regibo}, S. and {Huygen}, R. and {Royer}, P. and {Paproth}, C. and {Grie{\ss}bach}, D. and {Samadi}, R. and {Reese}, D.~R. and {Pertenais}, M. and {Grolleau}, E. and {Heller}, R. and {Niemi}, S.~M. and {Cabrera}, J. and {B{\"o}rner}, A. and {Aigrain}, S. and {McCormac}, J. and {Verhoeve}, P. and {Astier}, P. and {Kutrowski}, N. and {Vandenbussche}, B. and {Tkachenko}, A. and {Aerts}, C.},
        title = "{PlatoSim: an end-to-end PLATO camera simulator for modelling high-precision space-based photometry}",
      journal = {\aap},
         year = 2024,
        month = jan,
       volume = {681},
          eid = {A18},
        pages = {A18},
          doi = {10.1051/0004-6361/202346701},
archivePrefix = {arXiv},
       eprint = {2310.06985},
 primaryClass = {astro-ph.IM},
       adsurl = {https://ui.adsabs.harvard.edu/abs/2024A&A...681A..18J}
}

@ARTICLE{2022A&A...665A..11H,
       author = {{Heller}, Ren{\'e} and {Harre}, Jan-Vincent and {Samadi}, R{\'e}za},
        title = "{Transit least-squares survey. IV. Earth-like transiting planets expected from the PLATO mission}",
      journal = {\aap},
         year = 2022,
        month = sep,
       volume = {665},
          eid = {A11},
        pages = {A11},
          doi = {10.1051/0004-6361/202141640},
archivePrefix = {arXiv},
       eprint = {2206.02071},
 primaryClass = {astro-ph.EP},
       adsurl = {https://ui.adsabs.harvard.edu/abs/2022A&A...665A..11H}
}

@ARTICLE{2021A&A...653A..98M,
       author = {{Montalto}, M. and {Piotto}, G. and {Marrese}, P.~M. and {Nascimbeni}, V. and {Prisinzano}, L. and {Granata}, V. and {Marinoni}, S. and {Desidera}, S. and {Ortolani}, S. and {Aerts}, C. and {Alei}, E. and {Altavilla}, G. and {Benatti}, S. and {B{\"o}rner}, A. and {Cabrera}, J. and {Claudi}, R. and {Deleuil}, M. and {Fabrizio}, M. and {Gizon}, L. and {Goupil}, M.~J. and {Heras}, A.~M. and {Magrin}, D. and {Malavolta}, L. and {Mas-Hesse}, J.~M. and {Pagano}, I. and {Paproth}, C. and {Pertenais}, M. and {Pollacco}, D. and {Ragazzoni}, R. and {Ramsay}, G. and {Rauer}, H. and {Udry}, S.},
        title = "{The all-sky PLATO input catalogue}",
      journal = {\aap},
         year = 2021,
        month = sep,
       volume = {653},
          eid = {A98},
        pages = {A98},
          doi = {10.1051/0004-6361/202140717},
archivePrefix = {arXiv},
       eprint = {2108.13712},
 primaryClass = {astro-ph.EP},
       adsurl = {https://ui.adsabs.harvard.edu/abs/2021A&A...653A..98M}
}

@ARTICLE{2018A&A...616A...1G,
       author = {{Gaia Collaboration} and {Brown}, A.~G.~A. and {Vallenari}, A. and {Prusti}, T. and {de Bruijne}, J.~H.~J. and {Babusiaux}, C. and {Bailer-Jones}, C.~A.~L. and {Biermann}, M. and {Evans}, D.~W. and {Eyer}, L. and {Jansen}, F. and {Jordi}, C. and {Klioner}, S.~A. and {Lammers}, U. and {Lindegren}, L. and {Luri}, X. and {Mignard}, F. and {Panem}, C. and {Pourbaix}, D. and {Randich}, S. and {Sartoretti}, P. and {Siddiqui}, H.~I. and {Soubiran}, C. and {van Leeuwen}, F. and {Walton}, N.~A. and {Arenou}, F. and {Bastian}, U. and {Cropper}, M. and {Drimmel}, R. and {Katz}, D. and {Lattanzi}, M.~G. and {Bakker}, J. and {Cacciari}, C. and {Casta{\~n}eda}, J. and {Chaoul}, L. and {Cheek}, N. and {De Angeli}, F. and {Fabricius}, C. and {Guerra}, R. and {Holl}, B. and {Masana}, E. and {Messineo}, R. and {Mowlavi}, N. and {Nienartowicz}, K. and {Panuzzo}, P. and {Portell}, J. and {Riello}, M. and {Seabroke}, G.~M. and {Tanga}, P. and {Th{\'e}venin}, F. and {Gracia-Abril}, G. and {Comoretto}, G. and {Garcia-Reinaldos}, M. and {Teyssier}, D. and {Altmann}, M. and {Andrae}, R. and {Audard}, M. and {Bellas-Velidis}, I. and {Benson}, K. and {Berthier}, J. and {Blomme}, R. and {Burgess}, P. and {Busso}, G. and {Carry}, B. and {Cellino}, A. and {Clementini}, G. and {Clotet}, M. and {Creevey}, O. and {Davidson}, M. and {De Ridder}, J. and {Delchambre}, L. and {Dell'Oro}, A. and {Ducourant}, C. and {Fern{\'a}ndez-Hern{\'a}ndez}, J. and {Fouesneau}, M. and {Fr{\'e}mat}, Y. and {Galluccio}, L. and {Garc{\'\i}a-Torres}, M. and {Gonz{\'a}lez-N{\'u}{\~n}ez}, J. and {Gonz{\'a}lez-Vidal}, J.~J. and {Gosset}, E. and {Guy}, L.~P. and {Halbwachs}, J.-L. and {Hambly}, N.~C. and {Harrison}, D.~L. and {Hern{\'a}ndez}, J. and {Hestroffer}, D. and {Hodgkin}, S.~T. and {Hutton}, A. and {Jasniewicz}, G. and {Jean-Antoine-Piccolo}, A. and {Jordan}, S. and {Korn}, A.~J. and {Krone-Martins}, A. and {Lanzafame}, A.~C. and {Lebzelter}, T. and {L{\"o}ffler}, W. and {Manteiga}, M. and {Marrese}, P.~M. and {Mart{\'\i}n-Fleitas}, J.~M. and {Moitinho}, A. and {Mora}, A. and {Muinonen}, K. and {Osinde}, J. and {Pancino}, E. and {Pauwels}, T. and {Petit}, J.-M. and {Recio-Blanco}, A. and {Richards}, P.~J. and {Rimoldini}, L. and {Robin}, A.~C. and {Sarro}, L.~M. and {Siopis}, C. and {Smith}, M. and {Sozzetti}, A. and {S{\"u}veges}, M. and {Torra}, J. and {van Reeven}, W. and {Abbas}, U. and {Abreu Aramburu}, A. and {Accart}, S. and {Aerts}, C. and {Altavilla}, G. and {{\'A}lvarez}, M.~A. and {Alvarez}, R. and {Alves}, J. and {Anderson}, R.~I. and {Andrei}, A.~H. and {Anglada Varela}, E. and {Antiche}, E. and {Antoja}, T. and {Arcay}, B. and {Astraatmadja}, T.~L. and {Bach}, N. and {Baker}, S.~G. and {Balaguer-N{\'u}{\~n}ez}, L. and {Balm}, P. and {Barache}, C. and {Barata}, C. and {Barbato}, D. and {Barblan}, F. and {Barklem}, P.~S. and {Barrado}, D. and {Barros}, M. and {Barstow}, M.~A. and {Bartholom{\'e} Mu{\~n}oz}, S. and {Bassilana}, J.-L. and {Becciani}, U. and {Bellazzini}, M. and {Berihuete}, A. and {Bertone}, S. and {Bianchi}, L. and {Bienaym{\'e}}, O. and {Blanco-Cuaresma}, S. and {Boch}, T. and {Boeche}, C. and {Bombrun}, A. and {Borrachero}, R. and {Bossini}, D. and {Bouquillon}, S. and {Bourda}, G. and {Bragaglia}, A. and {Bramante}, L. and {Breddels}, M.~A. and {Bressan}, A. and {Brouillet}, N. and {Br{\"u}semeister}, T. and {Brugaletta}, E. and {Bucciarelli}, B. and {Burlacu}, A. and {Busonero}, D. and {Butkevich}, A.~G. and {Buzzi}, R. and {Caffau}, E. and {Cancelliere}, R. and {Cannizzaro}, G. and {Cantat-Gaudin}, T. and {Carballo}, R. and {Carlucci}, T. and {Carrasco}, J.~M. and {Casamiquela}, L. and {Castellani}, M. and {Castro-Ginard}, A. and {Charlot}, P. and {Chemin}, L. and {Chiavassa}, A. and {Cocozza}, G. and {Costigan}, G. and {Cowell}, S. and {Crifo}, F. and {Crosta}, M. and {Crowley}, C. and {Cuypers}, J. and {Dafonte}, C. and {Damerdji}, Y. and {Dapergolas}, A. and {David}, P. and {David}, M. and {de Laverny}, P. and {De Luise}, F.},
        title = "{Gaia Data Release 2. Summary of the contents and survey properties}",
      journal = {\aap},
         year = 2018,
        month = aug,
       volume = {616},
          eid = {A1},
        pages = {A1},
          doi = {10.1051/0004-6361/201833051},
archivePrefix = {arXiv},
       eprint = {1804.09365},
 primaryClass = {astro-ph.GA},
       adsurl = {https://ui.adsabs.harvard.edu/abs/2018A&A...616A...1G}
}

@ARTICLE{2022A&A...658A..31N,
       author = {{Nascimbeni}, V. and {Piotto}, G. and {B{\"o}rner}, A. and {Montalto}, M. and {Marrese}, P.~M. and {Cabrera}, J. and {Marinoni}, S. and {Aerts}, C. and {Altavilla}, G. and {Benatti}, S. and {Claudi}, R. and {Deleuil}, M. and {Desidera}, S. and {Fabrizio}, M. and {Gizon}, L. and {Goupil}, M.~J. and {Granata}, V. and {Heras}, A.~M. and {Magrin}, D. and {Malavolta}, L. and {Mas-Hesse}, J.~M. and {Ortolani}, S. and {Pagano}, I. and {Pollacco}, D. and {Prisinzano}, L. and {Ragazzoni}, R. and {Ramsay}, G. and {Rauer}, H. and {Udry}, S.},
        title = "{The PLATO field selection process. I. Identification and content of the long-pointing fields}",
      journal = {\aap},
         year = 2022,
        month = feb,
       volume = {658},
          eid = {A31},
        pages = {A31},
          doi = {10.1051/0004-6361/202142256},
archivePrefix = {arXiv},
       eprint = {2110.13924},
 primaryClass = {astro-ph.EP},
       adsurl = {https://ui.adsabs.harvard.edu/abs/2022A&A...658A..31N}
}

@ARTICLE{2025A&A...694A.313N,
       author = {{Nascimbeni}, V. and {Piotto}, G. and {Cabrera}, J. and {Montalto}, M. and {Marinoni}, S. and {Marrese}, P.~M. and {Aerts}, C. and {Altavilla}, G. and {Benatti}, S. and {B{\"o}rner}, A. and {Deleuil}, M. and {Desidera}, S. and {Gizon}, L. and {Goupil}, M.~J. and {Granata}, V. and {Heras}, A.~M. and {Magrin}, D. and {Malavolta}, L. and {Mas-Hesse}, J.~M. and {Osborn}, H.~P. and {Pagano}, I. and {Paproth}, C. and {Pollacco}, D. and {Prisinzano}, L. and {Ragazzoni}, R. and {Ramsay}, G. and {Rauer}, H. and {Tkachenko}, A. and {Udry}, S.},
        title = "{The PLATO field selection process: II. Characterization of LOPS2, the first long-pointing field}",
      journal = {\aap},
         year = 2025,
        month = feb,
       volume = {694},
          eid = {A313},
        pages = {A313},
          doi = {10.1051/0004-6361/202452325},
archivePrefix = {arXiv},
       eprint = {2501.07687},
 primaryClass = {astro-ph.EP},
       adsurl = {https://ui.adsabs.harvard.edu/abs/2025A&A...694A.313N}
}

@INPROCEEDINGS{2022SPIE12180E..1EK,
       author = {{Koncz}, Alexander and {Eigm{\"u}ller}, Philipp and {Michaelis}, Harald and {Rauer}, Heike and {Wolter}, David and {Tschentscher}, Matthias and {Weisse}, Stefan and {Vasiliou}, Konstantinos and {Tomecki}, Daniel and {Rufini}, Sergio and {Cacovean}, Andrei and {Althaus}, Christian and {Jung}, Boris and {Mueller}, Uwe and {Korsitzky}, Hartmut and {Terzer}, Ronny and {Manthey}, Kristian and {Grott}, Matthias and {Ligus}, Jan and {Hviid}, Stubbe and {Cara}, Christophe and {Lavanant}, Tony and {Fontigni{\`e}}, Jean and {Huynh}, Dat and {Nico}, Francois and {Niemi}, Sami},
        title = "{PLATO fast front end electronics (F-FEE): performance results of the engineering model}",
    booktitle = {Space Telescopes and Instrumentation 2022: Optical, Infrared, and Millimeter Wave},
         year = 2022,
       editor = {{Coyle}, Laura E. and {Matsuura}, Shuji and {Perrin}, Marshall D.},
       series = {Society of Photo-Optical Instrumentation Engineers (SPIE) Conference Series},
       volume = {12180},
        month = aug,
          eid = {121801E},
        pages = {121801E},
          doi = {10.1117/12.2629953},
       adsurl = {https://ui.adsabs.harvard.edu/abs/2022SPIE12180E..1EK}
}

@ARTICLE{2026AJ....171...14B,
       author = {{Bowling}, Sam and {Lieu}, R. and {Griessbach}, D.~G. and {Jannsen}, N. and {Cabrera}, J. and {Paproth}, C. and {Rauer}, H. and {Heller}, R. and {Jiang}, C.},
        title = "{The Effect of Guide Star Variability on PLATO Pointing Stability}",
      journal = {\aj},
         year = 2026,
        month = jan,
       volume = {171},
       number = {1},
          eid = {14},
        pages = {14},
          doi = {10.3847/1538-3881/ae196e},
       adsurl = {https://ui.adsabs.harvard.edu/abs/2026AJ....171...14B}
}

@TECHREPORT{2010Christiansen,
       author = {{Christiansen}, Jessie and {Machalek}, Pavel},
        title = "{Kepler Data Release 7 Notes}",
      journal = {KSCI-19047-001},
         year = "2010",
        month = "Sep",
        pages = {62}
}

@INPROCEEDINGS{2021SPIE11852E..3HG,
       author = {{Grie{\ss}bach}, Denis and {Witteck}, Ulrike and {Paproth}, Carsten},
        title = "{The fine guidance system of the PLATO mission}",
    booktitle = {International Conference on Space Optics {\textemdash} ICSO 2020},
         year = 2021,
       editor = {{Cugny}, Bruno and {Sodnik}, Zoran and {Karafolas}, Nikos},
       series = {Society of Photo-Optical Instrumentation Engineers (SPIE) Conference Series},
       volume = {11852},
        month = jun,
          eid = {118523H},
        pages = {118523H},
          doi = {10.1117/12.2599604},
       adsurl = {https://ui.adsabs.harvard.edu/abs/2021SPIE11852E..3HG}
}

@ARTICLE{2021A&A...653A.146G,
       author = {{Gurgenashvili}, E. and {Zaqarashvili}, T.~V. and {Kukhianidze}, V. and {Reiners}, A. and {Oliver}, R. and {Lanza}, A.~F. and {Reinhold}, T.},
        title = "{Rieger-type periodicity in the total irradiance of the Sun as a star during solar cycles 23-24}",
      journal = {\aap},
         year = 2021,
        month = sep,
       volume = {653},
          eid = {A146},
        pages = {A146},
          doi = {10.1051/0004-6361/202141370},
       adsurl = {https://ui.adsabs.harvard.edu/abs/2021A&A...653A.146G}
}

@ARTICLE{2026A&A...707A...2G,
       author = {{Guti{\'e}rrez-Canales}, F. and {Samadi}, R. and {Birch}, A. and {Cabrera}, J. and {Damiani}, C. and {Guterman}, P. and {Paproth}, C. and {Pertenais}, M. and {Santerne}, A.},
        title = "{Detecting false positives with PLATO using double-aperture photometry and centroid shifts}",
      journal = {\aap},
         year = 2026,
        month = feb,
       volume = {707},
          eid = {A2},
        pages = {A2},
          doi = {10.1051/0004-6361/202453294},
archivePrefix = {arXiv},
       eprint = {2512.18844},
 primaryClass = {astro-ph.EP},
       adsurl = {https://ui.adsabs.harvard.edu/abs/2026A&A...707A...2G}
}

@ARTICLE{2013A&A...557L..10N,
       author = {{Nielsen}, M.~B. and {Gizon}, L. and {Schunker}, H. and {Karoff}, C.},
        title = "{Rotation periods of 12 000 main-sequence Kepler stars: Dependence on stellar spectral type and comparison with v sin i observations}",
      journal = {\aap},
         year = 2013,
        month = sep,
       volume = {557},
          eid = {L10},
        pages = {L10},
          doi = {10.1051/0004-6361/201321912},
archivePrefix = {arXiv},
       eprint = {1305.5721},
 primaryClass = {astro-ph.SR},
       adsurl = {https://ui.adsabs.harvard.edu/abs/2013A&A...557L..10N}
}

@ARTICLE{1972ApJ...171..565S,
       author = {{Skumanich}, A.},
        title = "{Time Scales for Ca II Emission Decay, Rotational Braking, and Lithium Depletion}",
      journal = {\apj},
         year = 1972,
        month = feb,
       volume = {171},
        pages = {565},
          doi = {10.1086/151310},
       adsurl = {https://ui.adsabs.harvard.edu/abs/1972ApJ...171..565S}
}

@ARTICLE{2007ApJ...669.1167B,
       author = {{Barnes}, Sydney A.},
        title = "{Ages for Illustrative Field Stars Using Gyrochronology: Viability, Limitations, and Errors}",
      journal = {\apj},
         year = 2007,
        month = nov,
       volume = {669},
       number = {2},
        pages = {1167-1189},
          doi = {10.1086/519295},
archivePrefix = {arXiv},
       eprint = {0704.3068},
 primaryClass = {astro-ph},
       adsurl = {https://ui.adsabs.harvard.edu/abs/2007ApJ...669.1167B}
}

@ARTICLE{2003ApJ...586..464B,
       author = {{Barnes}, Sydney A.},
        title = "{On the Rotational Evolution of Solar- and Late-Type Stars, Its Magnetic Origins, and the Possibility of Stellar Gyrochronology}",
      journal = {\apj},
         year = 2003,
        month = mar,
       volume = {586},
       number = {1},
        pages = {464-479},
          doi = {10.1086/367639},
archivePrefix = {arXiv},
       eprint = {astro-ph/0303631},
 primaryClass = {astro-ph},
       adsurl = {https://ui.adsabs.harvard.edu/abs/2003ApJ...586..464B}
}

\end{document}